\documentclass[aps,prc,twocolumn,showpacs,amsmath,amssymb,nofootinbib]{revtex4-1}

\usepackage{graphicx,latexsym,amssymb,epsfig}
\usepackage{color}
\usepackage{times}
\usepackage{diagbox}
\usepackage{amsmath}
\usepackage[normalem]{ulem}
\usepackage{inputenc}
\usepackage{bm}
\usepackage{multirow}
\usepackage{makecell}
\usepackage{float}
\usepackage{url}
\usepackage{natbib}
\usepackage[colorlinks=true,citecolor=blue,urlcolor=blue,linkcolor=red]{hyperref}
\usepackage{tikz,xcolor,hyperref}
\usepackage{subfigure}
\usepackage{tabularx}
\usepackage[figuresright]{rotating}

\newcommand{\Msun}{\,{M}_{\odot}}
\def\fm3{\;\text{fm}^{-3}}

\definecolor{lime}{HTML}{A6CE39}
\DeclareRobustCommand{\orcidicon}{%
	\begin{tikzpicture}
		\draw[lime, fill=lime] (0,0)
		circle [radius=0.17]
		node[white] {{\fontfamily{qag}\selectfont \tiny ID}};
		\draw[white, fill=white] (-0.0625,0.095)
		circle [radius=0.008];
	\end{tikzpicture}
	\hspace{-2mm}
}
\foreach \x in {A, ..., Z}{
	\expandafter\xdef\csname orcid\x\endcsname{\noexpand\href{https://orcid.org/\csname orcidauthor\x\endcsname}{\noexpand\orcidicon}}
}

\begin{document}

\title{Nuclear pasta in hot neutron-star matter and proto-neutron stars}

\author{Jian Zhou$^{1}$}

\author{Junbo Pang$^{1}$}

\author{Hong Shen{\orcidA{}}$^{1}$}
\email{shennankai@gmail.com}

\author{Jinniu Hu{\orcidB{}}$^{1}$}
\email{hujinniu@nankai.edu.cn}

\affiliation{\it
$^{1}$ School of Physics, Nankai University, Tianjin 300071, China \\
}


\begin{abstract}
We investigate nuclear pasta phases appearing in hot neutron-star matter based 
on the compressible liquid-drop model, where the matter consists 
of a dense liquid phase and a dilute gas phase separated by a sharp interface.
The surface tension is calculated self-consistently 
from the Thomas--Fermi approximation, and it depends on temperature and isospin asymmetry.
We employ relativistic mean-field (RMF) models with different symmetry energy slopes 
to describe nuclear interactions. It is found that the RMF models with smaller symmetry 
energy slopes result in various pasta phases at low temperatures, whereas those 
with larger symmetry energy slopes yield only the droplet configuration 
up to the crust-core transition density.
We examine the occurrence and influence of pasta phases in proto-neutron 
stars with a constant entropy per baryon. These pasta phases may occur in the inner crust
and play an important role in the thermal evolution of the star.
\end{abstract}

\maketitle
\section{Introduction}
\label{sec:1}

A proto-neutron star (PNS) is the compact object formed after the gravitational collapse 
of a massive stellar core~\cite{fish10,burr13,yama24}. 
It represents a fascinating transitional phase in stellar 
evolution, preceding the formation of cold neutron stars.
After its formation in a core-collapse supernova, the PNS cools through deleptonization
and neutrino emission. Within tens of seconds, neutrinos can escape freely from the PNS 
as the entropy steadily decreases~\cite{sumi19,naka19}.
During the cooling process, the PNS contracts rapidly and heavy nuclei may appear in the 
surface region~\cite{naka18,pote21}. The formation of heavy nuclei indicates a transition 
from uniform matter to a nonuniform crust.
This stage is characterized by extreme physical conditions that challenge our understanding 
of nuclear physics and the thermal evolution of compact stars.

The structure and composition of the PNS are studied using the equation of state (EOS) of 
hot nuclear matter in $\beta$ equilibrium, referred to as hot neutron-star matter.
The EOS plays a decisive role in understanding astrophysical phenomena, such as 
core-collapse supernova explosions, PNS cooling, and various properties of neutron
stars~\cite{cham08,latt16,oert17}. 
Over the past decades, astronomical observational technologies 
have achieved significant advances, which have greatly improved our understanding 
of neutron-star properties~\cite{koeh25,chat24,asce24}. 
These observations provide a wealth of information for constraining physics of matter 
under extreme conditions. The first direct detection of gravitational waves from a binary black 
hole merger, known as GW150914~\cite{abbo16}, launched a new era of gravitational-wave astronomy.
The observations of compact binary mergers by the LIGO and Virgo
detectors have provided valuable constraints on the EOS and its underlying theoretical framework.
It is well known that the tidal deformability of neutron stars can be inferred from 
gravitational-wave observations, although a large uncertainty remains from both the measurements 
and theoretical analyses. 
The milestone gravitational wave event GW170817~\cite{abbo17,abbo18} from a binary neutron-star 
merger provided an estimate of the tidal deformability and constrained the radius for canonical 
neutron stars with masses around 1.4$\Msun$~\cite{fatt18,most18,land20,chat20}.
Furthermore, several precise measurements of massive pulsars~\cite{demo10,arzo18,anto13,fons21},
require the maximum neutron-star mass to be larger than 2$\, M_\odot$,
which provides a stringent constraint on the EOS of neutron stars.
The recent observations by NICER (Neutron Star Interior Composition Explorer)
for PSR J0030+0451~\cite{mill19,rile19} and PSR J0740+6620~\cite{mill21,rile21}
have provided simultaneous measurements of the mass and radius of neutron stars,
which offer further constraints on the EOS of dense matter.
The nuclear many-body approach employed in studies of the EOS should be compatible with 
current constraints from astrophysical observations and with various properties of finite nuclei.

In this work, we employ the relativistic mean-field (RMF) approach to describe nuclear interactions.
Currently, the RMF approach is recognized as a specific realization of covariant density functional 
theory, providing a consistent relativistic framework applicable to nuclear physics and 
astrophysics.
Over the past decades, a number of RMF parameter sets have been developed that provide a satisfactory 
description of both finite nuclei and various astrophysical objects 
(see, e.g., Refs.~\cite{dutr14,oert17,miya25} for reviews).
Generally, RMF parametrizations were obtained by fitting experimental data for selected nuclei, 
including their binding energies and charge radii, while some nuclear matter properties were also 
considered in the fitting procedure. Consequently, most of the well-known RMF parametrizations 
provide reasonable descriptions of finite nuclei and similar saturation properties for symmetric 
nuclear matter. However, large uncertainties still remain in asymmetric nuclear matter, especially 
the symmetry energy and its density dependence, due to the lack of well-measured isovector 
observables in the fitting procedure. 
In recent years, theoretical uncertainty estimates have attracted considerable attention
in order to assess the reliability of extrapolations~\cite{PRA11}.
Unfortunately, most of the successful RMF models presented in Ref.~\cite{dutr14}
did not consider theoretical uncertainties when they were calibrated against selected
experimental data of finite nuclei. 
In Ref.~\cite{chen14}, the authors proposed the FSUGold2 parameter set within the RMF framework,
in which statistical uncertainties of various physical observables 
were quantified by using a Gaussian approximation to the likelihood function. 
The uncertainty estimates obtained with the FSUGold2 parametrization could be considered as 
typical values for understanding theoretical uncertainties in RMF models.
With a proper calibration procedure including both experimental data of finite nuclei
and a maximum neutron-star mass of $M_\textrm{TOV}\sim 2.1~M_\odot$, the FSUGold2 parametrization 
yielded predictions for neutron-star observables with estimated uncertainties of
$M_\textrm{TOV}=2.07 \pm 0.02~M_\odot$ and $R_{1.4} =14.42 \pm 0.26$~km.
Moreover, beyond the statistical uncertainties discussed in Ref.~\cite{chen14},
RMF models contain a variety of systematic uncertainties that are difficult to estimate
quantitatively~\cite{doba14}. One may assess systematic uncertainties of RMF models
by comparing different parametrizations~\cite{yang20}.

In the present work, we employ two pairs of RMF parametrizations based on the TM1~\cite{suga94} 
and IUFSU~\cite{fatt10} models to investigate the influence of nuclear symmetry energy and its slope
on neutron-star matter.
The first pair consists of the original TM1 parametrization proposed in Ref.~\cite{suga94} 
and its extended version TM1e~\cite{bao14b,ji19,shen20}. 
It is noteworthy that the TM1 and TM1e models have the same isoscalar properties but
exhibit significantly different behaviors regarding nuclear symmetry energy. 
The corresponding slope parameters of the symmetry energy are $L=40$ MeV for TM1e and $L=110.8$ MeV for TM1. 
Similarly, the second pair consists of the original IUFSU parametrization 
with $L=47.2$ MeV~\cite{fatt10} and a modified IUFSU parametrization with $L=110$ MeV~\cite{bao14b}. 
It is well known that the slope parameter $L$ plays a crucial role in determining 
the neutron-skin thickness of finite nuclei and is also strongly correlated with neutron-star 
radii and crust structure~\cite{horo01,bao15,ji19}.
In our previous work~\cite{bao14b}, we simultaneously adjusted two parameters associated 
with the $\rho$ meson in an RMF model to generate modified models with a given slope 
parameter $L$ at saturation density while maintaining the same symmetry energy as the 
original RMF model at a density of 0.11 fm$^{-3}$.
The choice of the fixed density $n_{\rm fix}=0.11\,\rm{fm}^{-3}$ for the symmetry energy 
is motivated by the fact that this value is approximately the average density of finite nuclei. 
Thus, imposing the same symmetry energy at $n_{\rm fix}=0.11\,\rm{fm}^{-3}$ can result in 
nearly unchanged binding energies of $^{208}$Pb for different values 
of $L$ (see Fig.~1 of Ref.~\cite{bao14b}).
It is noteworthy that the modified and original models have identical isoscalar
properties and the same symmetry energy at $0.11\, \rm{fm}^{-3}$, so that 
both models can provide very similar descriptions of stable nuclei.
However, due to their different slope parameters $L$,
they yield different neutron-star radii and different behaviors of asymmetric nuclear matter.
By comparing the results from a pair of RMF models with different values of $L$, 
we can examine the effects arising solely from the behavior of the symmetry energy.
In addition, the TM1e and TM1 models have been used to construct general EOSs 
for astrophysical simulations, such as core-collapse supernovae and binary 
neutron-star mergers~\cite{shen20,lisy25a}, and to explore the properties of pasta 
phases in supernova matter~\cite{ji20}.

At low temperatures and subsaturation densities, heavy nuclei are formed in nonuniform matter
to reduce the free energy of the system~\cite{rave83,okam13,shen11}. 
As the density increases close to the crust-core transition, nonspherical pasta phases may appear, 
arising from the competition between surface and Coulomb energies~\cite{bao15}.
As a result, the stable nuclear shape in nonuniform matter may change from spherical droplets to rods,
slabs, tubes, and bubbles with increasing baryon density. 
Possibly, nuclear pasta phases exist both in core-collapse supernova matter with fixed proton fraction 
and in the inner crust of neutron stars, where the matter is neutron rich and in $\beta$ equilibrium.
Over the past decades, nuclear pasta phases have been studied using various methods, 
such as the liquid-drop model~\cite{wata00,ji20} 
and the Thomas-Fermi approximation~\cite{oyam07,avan10,gril12,pais15,xia22}.
It is noteworthy that nuclear symmetry energy and its density dependence significantly
affect the pasta phase structure and determine its density region~\cite{oyam07,gril12,bao15,ji20}.
In addition, the structure and properties of nuclear pasta can alter thermal conductivity 
and neutrino transport in a PNS, thereby influencing the PNS cooling process. 
It is of great interest and importance to investigate nuclear pasta phases in hot neutron-star matter.

We have two aims in this article. The first is to investigate the properties
of pasta phases occurring in hot neutron-star matter.
Specifically, we examine the effects of nuclear symmetry energy and its density 
dependence by comparing the results obtained using two pairs of RMF parametrizations
based on the TM1 and IUFSU models, with different values of the slope parameter $L$ 
within each pair.
The second is to explore the impact of pasta phases on PNS properties, 
employing a constant entropy approach.
To describe nuclear pasta phases at finite temperatures, we adopt the compressible 
liquid-drop (CLD) method with RMF models for nuclear interactions.
Generally, the Wigner-Seitz approximation is employed to simplify the calculations, 
in which the whole space is divided into equivalent cells of specific geometric symmetry.
Within each Wigner-Seitz cell, a dense liquid coexists with a dilute gas of free nucleons 
and $\alpha$ particles, with the two phases separated by a sharp interface.
The equilibrium conditions for two coexisting phases are derived by minimizing the total 
free energy, which includes surface and Coulomb contributions. These conditions, which 
incorporate finite-size effects, differ from the Gibbs conditions for phase equilibrium.
When finite-size effects are neglected, the CLD method reduces to a bulk matter calculation 
satisfying the Gibbs equilibrium conditions.
In principle, the Thomas-Fermi approximation provides a more realistic description of 
pasta phases than the CLD method, since it yields the density profile within the Wigner-Seitz cell 
self-consistently. A detailed comparison between the CLD method and the Thomas-Fermi approximation
for nonuniform matter at zero temperature was reported in Ref.~\cite{bao14b},
which indicates that both methods yield similar results for neutron-star crusts.
For calculations of hot neutron-star matter, the Thomas-Fermi approximation is
computationally more expensive than the CLD method.
Therefore, in the present work, we adopt the CLD method to explore the properties 
of pasta phases in hot neutron-star matter.
  
The surface tension $\tau$ plays a key role in determining the structure of pasta 
phases~\cite{avan10,ji20}. Generally, $\tau$ depends on the temperature $T$ and 
the proton fraction of the dense liquid phase $Y_p^L$~\cite{schn17,bao16}.
The surface tension can be derived by using the Thomas-Fermi approximation for 
a one-dimensional nuclear system consisting of protons and neutrons~\cite{avan10,bao14a}.
In our previous works~\cite{bao16,ji20}, we calculated the surface tension $\tau$ at each
given $T$ and $Y_p^L$, a computationally expensive procedure.
For simplicity, in the present calculation, we employ a parameterized surface tension $\tau$, 
with parameters fitted to the results of the Thomas-Fermi approximation.

This article is organized as follows. In Sec.~\ref{sec:2},
we briefly review the RMF approach and the CLD method used for describing nuclear 
pasta phases in hot neutron-star matter.
In Sec.~\ref{sec:3}, we present the results for nuclear pasta in hot neutron-star 
matter and discuss their occurrence and influence in a PNS.
The conclusions are given in Sec.~\ref{sec:4}.

\section{Formalism}
\label{sec:2}

We investigate nuclear pasta phases appearing in hot neutron-star matter based on the 
CLD method~\cite{bao16,ji20}, where the RMF model is adopted for the nuclear interaction~\cite{bao14b}.
For the self-containment of this article, we provide a brief description of the RMF model 
and the CLD method. 

\subsection{RMF model}
\label{sec:2.1}

In the RMF approach, nucleons interact via the exchange of mesons, including 
the isoscalar-scalar meson $\sigma$, isoscalar-vector meson $\omega$, and isovector-vector meson $\rho$.
The Lagrangian density of neutron-star matter consisting of neutrons, protons, and electrons
can be written as
\begin{eqnarray}
\label{eq:LRMF}
\mathcal{L} & = & \sum_{i=p,n}\bar{\psi}_i \left[
i\gamma_{\mu}\partial^{\mu}-\left(M+g_{\sigma}\sigma\right)
 \right. \notag \\  & & \left.
-\gamma_{\mu} \left(g_{\omega}\omega^{\mu}
+\frac{g_{\rho}}{2} \tau_a\rho^{a\mu}\right) \right]\psi_i  \notag \\
&& +\frac{1}{2}\partial_{\mu}\sigma\partial^{\mu}\sigma -\frac{1}{2}%
m^2_{\sigma}\sigma^2-\frac{1}{3}g_{2}\sigma^{3} -\frac{1}{4}g_{3}\sigma^{4}
\notag \\
&& -\frac{1}{4}W_{\mu\nu}W^{\mu\nu} +\frac{1}{2}m^2_{\omega}\omega_{\mu}%
\omega^{\mu} +\frac{1}{4}c_{3}\left(\omega_{\mu}\omega^{\mu}\right)^2  \notag
\\
&& -\frac{1}{4}R^a_{\mu\nu}R^{a\mu\nu} +\frac{1}{2}m^2_{\rho}\rho^a_{\mu}\rho^{a\mu}
+\Lambda_{\rm{V}} \left(g_{\omega}^2
\omega_{\mu}\omega^{\mu}\right)
\left(g_{\rho}^2\rho^a_{\mu}\rho^{a\mu}\right) \notag\\
&& +\bar{\psi}_e(i\gamma_\mu\partial^\mu-m_e)\psi_e,
\end{eqnarray}
where $W^{\mu\nu}$ and $R^{a\mu\nu}$ denote the antisymmetric field tensors for $\omega^{\mu}$ 
and $\rho^{a\mu}$, respectively. The isospin Pauli matrices are denoted by $\tau_a$, with the third component $\tau_3=1$ for protons and $\tau_3=-1$ for neutrons. With the mean-field approximation, the meson fields are treated 
as classical fields, replacing the field operators with their expectation values.
The nonvanishing expectation values of meson fields are denoted as 
$\sigma =\left\langle \sigma \right\rangle$, $\omega =\left\langle \omega^{0}\right\rangle$,
and $\rho =\left\langle \rho^{30} \right\rangle$.
In homogeneous neutron-star matter at finite temperature, it is straightforward to derive the 
equations of motion for nucleons and mesons~\cite{bao16}. 
These coupled equations can be solved self-consistently. 
The thermodynamic quantities of interest include the energy density given by
\begin{eqnarray}
\epsilon&=&\sum_{i=p,n,e}\frac{1}{\pi^{2}}\int_{0}^{\infty} dk\:k^{2}\:\sqrt{k^{2}+m_i^{*2}}
\:(f_{i+}^{k}\:+f_{i-}^{k})  \notag\\
&&+\frac{1}{2}m_{\sigma}^{2}\sigma^{2}+\frac{1}{3}g_{2}\sigma^{3}+\frac{1}{4}g_{3}\sigma^{4}
 +\frac{1}{2}m_{\omega}^{2}\omega^{2}  \notag\\
&&+\frac{3}{4}c_{3}\omega^{4}+\frac{1}{2}m_{\rho}^{2}\rho^{2}
+3\Lambda_{\rm{V}}(g_{\omega}^{2}\omega^{2})(g_{\rho}^{2}\rho^{2}),
\label{eq:energy}
\end{eqnarray}
the entropy density written as
\begin{eqnarray}
s&=&-\sum_{i=p,n,e}\frac{1}{\pi^{2}}\int_{0}^{\infty} dk\:k^{2}
\left[f_{i+}^{k}\ln f_{i+}^{k} \right.  \notag\\
&&+(1-f_{i+}^{k})\ln(1-f_{i+}^{k})+f_{i-}^{k}\ln f_{i-}^{k}  \notag\\
&&\left. +(1-f_{i-}^{k})\ln(1-f_{i-}^{k})\right],
\label{eq:shp}
\end{eqnarray}
and the pressure given by
\begin{eqnarray}
P&=&\sum_{i=p,n,e}\frac{1}{3\pi^{2}}\int_{0}^{\infty} dk\:\frac{k^4}{\sqrt{k^2+m_i^{*2}}}
\:(f_{i+}^{k}\:+f_{i-}^{k})   \notag\\
&& -\frac{1}{2}m^2_{\sigma}{\sigma}^2-\frac{1}{3}{g_2}{\sigma}^3
  - \frac{1}{4}{g_3}{\sigma}^4  + \frac{1}{2}m^2_{\omega}{\omega}^2   \notag\\
&& +\frac{1}{4}{c_3}{\omega}^4
  + \frac{1}{2}m^2_{\rho}{\rho}^2
  + \Lambda_{\rm{V}}\left(g^2_{\omega}{\omega}^2\right)
  \left(g^2_{\rho}{\rho}^2\right).
\label{eq:php}
\end{eqnarray}
The free energy density is calculated from
\begin{eqnarray}
f=\epsilon-Ts.
\label{eq:fhp}
\end{eqnarray}
Here, $m_p^*=m_n^*=M+g_\sigma\sigma$ is the effective nucleon mass, while $m_e^*=m_e$ denotes the electron mass. 
$f_{i+}^k$ and $f_{i-}^k$ represent the occupation probabilities of particle and antiparticle at momentum $k$, which are given by the Fermi-Dirac distribution,
\begin{equation}
f_{i\pm}^{k}=\left\{1\:+\:\exp\:\left[\left(\sqrt{k^{2}+m_i^{*2}}\:\mp\:\nu_{i}\right)/T\right]\right\}^{-1} ,
\end{equation}
where $\nu_i$ denotes the kinetic part of the chemical potential $\mu_i$. 
For protons ($i=p$) and neutrons ($i=n$), $\nu_i=\mu_i-g_\omega\omega-\frac{g_\rho}{2}\tau_{3i}\rho$, 
whereas for electrons, $\nu_e=\mu_e$. 
The number density of species $i$ is calculated from 
\begin{eqnarray}
n_i = \frac{1}{\pi^2} \int_0^\infty \:dk \:k^2 \left(f_{i+}^k - f_{i-}^k\right).
\end{eqnarray}

\begin{table*}[tbp]
\caption{Coupling constants of the TM1e, TM1, and IUFSU models.}
\begin{center}
\begin{tabular}{lcccccccccccc}
\hline\hline
Model & $g_\sigma$  & $g_\omega$ & $g_\rho$ & $g_{2}$ (fm$^{-1}$) & $g_{3}$ & $c_{3}$ & $\Lambda_{\rm{V}}$ \\
\hline
TM1e $(L=40)$  & 10.0289     & 12.6139    & 13.9714  & $-$7.2325        &0.6183   & 71.3075 & 0.0429  \\
\hline
TM1 $(L=110.8)$  & 10.0289     & 12.6139    &  9.2644  & $-$7.2325        &0.6183   & 71.3075 & 0.0000  \\
\hline
IUFSU $(L=47.2)$  & 9.9713     & 13.0321    &  13.5899  & $-$8.4929        &0.4877   & 144.2195 & 0.0460  \\
\hline
IUFSU $(L=110)$  & 9.9713     & 13.0321    &  8.8192  & $-$8.4929        &0.4877   & 144.2195 & 0.0011  \\
\hline\hline
\end{tabular}
\label{tab:1}
\end{center}
\end{table*}

\begin{figure}[htbp]
\begin{center}
\includegraphics[clip,width=8.6 cm]{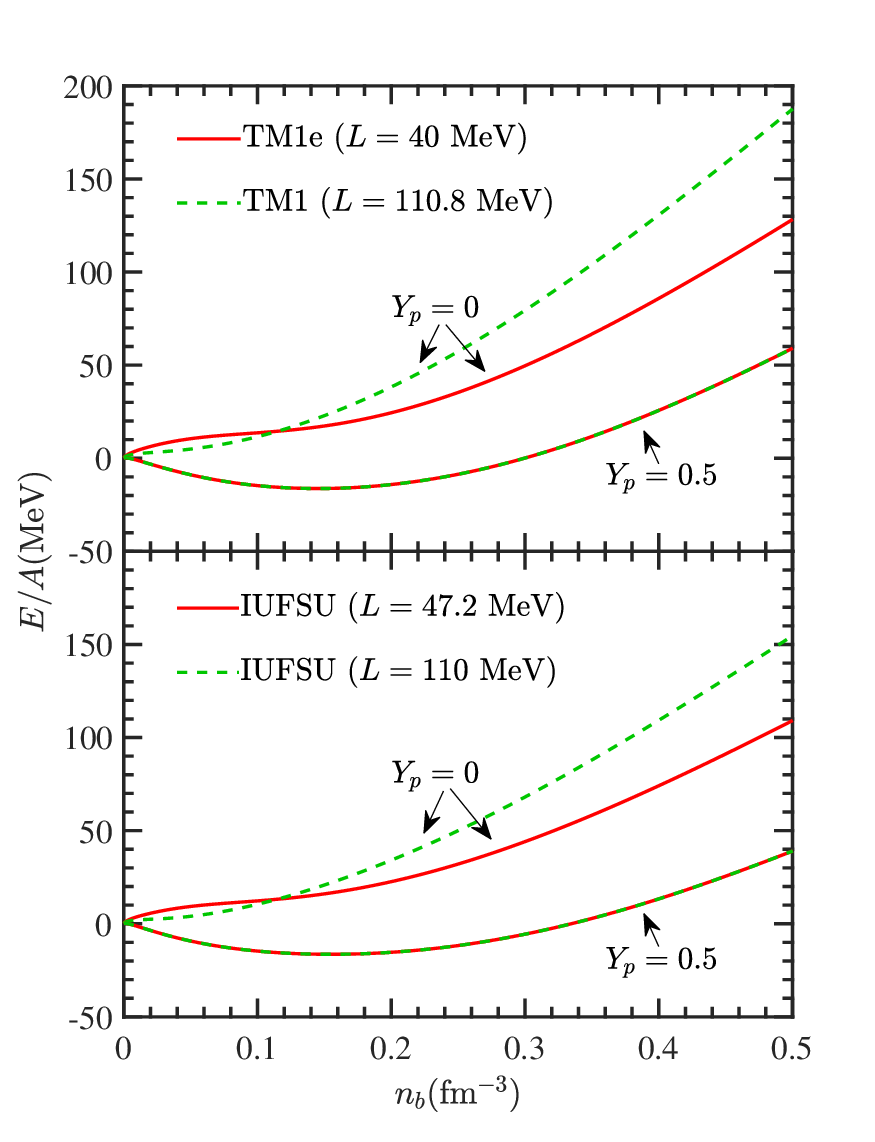}
\caption{Energy per baryon $E/A$ of symmetric nuclear matter and pure neutron matter as a function 
of the baryon number density $n_b$ in the TM1e and TM1 models (upper panel) 
and in the IUFSU models (lower panel).}
\label{fig:EA}
\end{center}
\end{figure}

\begin{figure}[!h]
\begin{center}
\includegraphics[clip,width=8.6 cm]{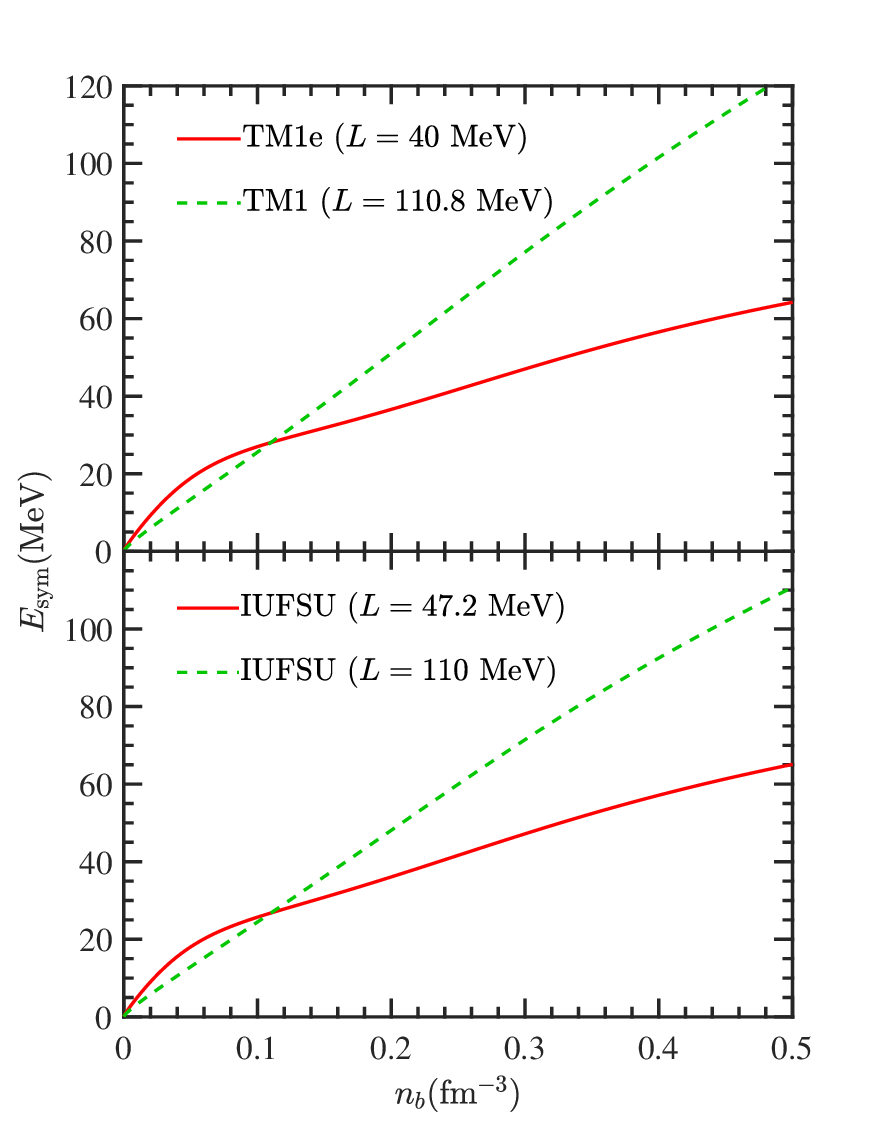}
\caption{Symmetry energy $E_{\rm sym}$ as a function of the baryon number density $n_b$ 
in the TM1e and TM1 models (upper panel) and in the IUFSU models (lower panel).}
\label{fig:Esym}
\end{center}
\end{figure}

\begin{figure*}[htbp]
	\begin{center}
		\includegraphics[clip,width=13.0 cm]{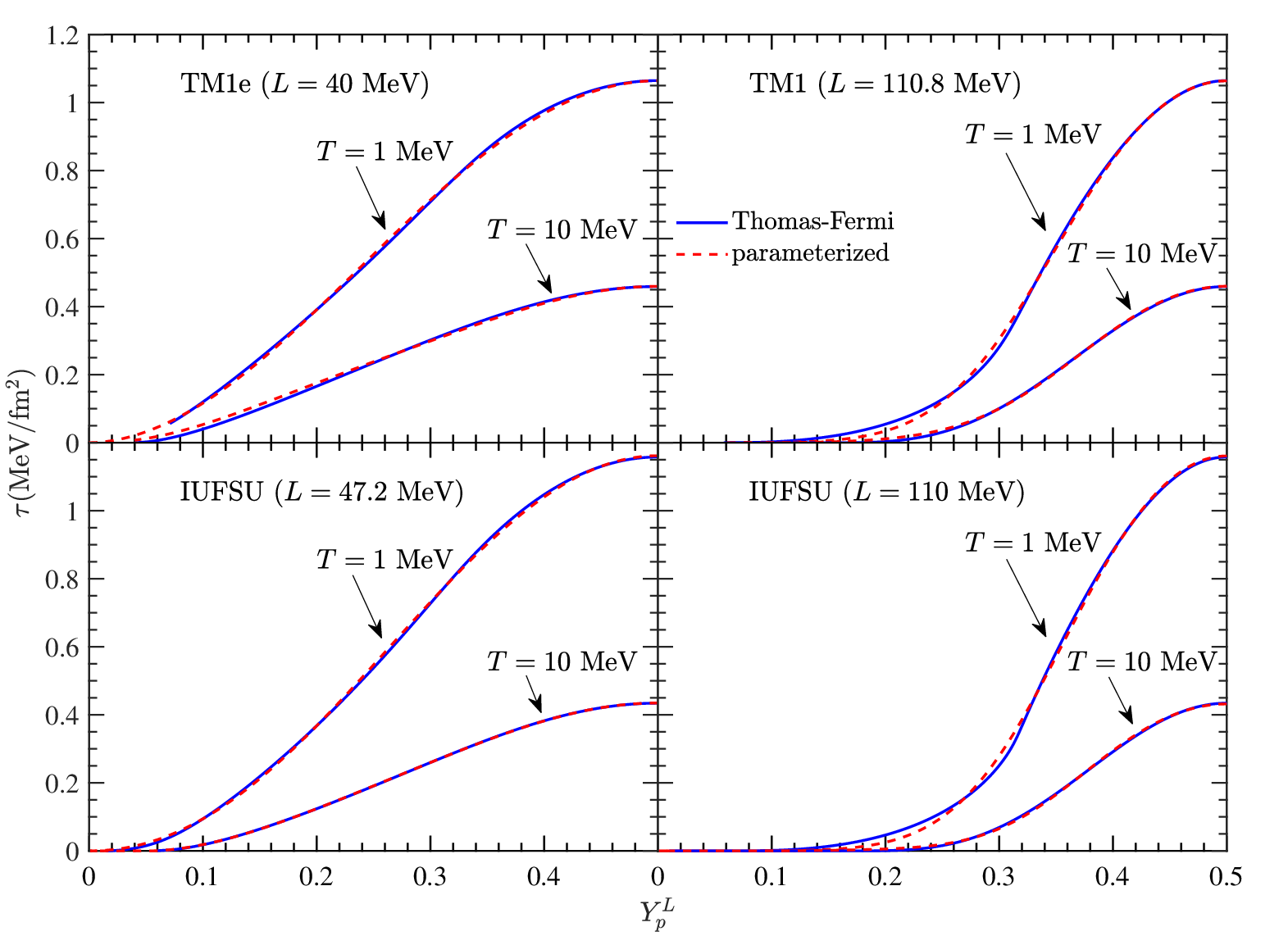}
		\caption{
            Surface tension $\tau$ as a function of the proton fraction in the dense liquid phase $Y_p^L$
            for $T=1$ and $10$ MeV. The parametrized results (red dashed lines) are compared with
            those from the Thomas-Fermi approximation (blue solid lines).} 
		\label{fig:suft}
	\end{center}
\end{figure*}

The RMF models with various parametrizations have been widely and successfully used to describe 
infinite nuclear matter, finite nuclei, and stellar matter in extreme isospin conditions~\cite{dutr14,miya25}.
In this study, we investigate the effects of the symmetry energy and its slope 
on nuclear pasta in hot neutron-star matter using two pairs of RMF parametrizations based 
on the TM1~\cite{suga94} and IUFSU~\cite{fatt10} models.
The first pair consists of the original TM1 model with a large slope parameter $L=110.8$ MeV
and its extended version TM1e with $L=40$ MeV.
The original TM1 model, proposed by Sugahara and Toki~\cite{suga94}, provides a satisfactory description 
of finite nuclei, including unstable nuclei, and has also been widely used to construct the EOS for 
supernova simulations~\cite{shen98b,shen11,oert17}. 
An extended TM1 model, referred to as the TM1e model, was proposed by~\citet{bao14b} 
and subsequently applied to construct the EOS table for astrophysical simulations~\cite{shen20}.
Similarly, the second pair consists of the original IUFSU model 
with $L=47.2$ MeV~\cite{fatt10} and a modified IUFSU model with $L=110$ MeV~\cite{bao14b}.
For completeness, we present the coupling constants of these models in Table~\ref{tab:1}.
It is noteworthy that, while the TM1e and original TM1 models share the same isoscalar properties, 
they exhibit different behaviors in the symmetry energy. 
Compared to the original TM1 model, an additional $\omega$--$\rho$ coupling term characterized by
the coupling $\Lambda_{\rm{V}}$ is introduced in the TM1e model, which plays a crucial role 
in determining the density dependence of the symmetry energy~\cite{bao14a,bao14b}.
In the TM1e model, at saturation density, the symmetry energy $E_{\text{sym}}=31.38$ MeV and 
the slope parameter $L=40$ MeV are obtained, which fall well within the constraints from various
observations~\cite{oert17}. In contrast, the original TM1 model yields 
$E_{\text{sym}}=36.89$ MeV and $L=110.8$ MeV, which predict rather large neutron-star radii.
Similar correlations between the slope parameter $L$ and the symmetry energy $E_{\text{sym}}$ 
are also observed for the pair of IUFSU models.
By comparing the results from a pair of RMF models with different values of $L$, 
we can examine the effects arising solely from the behavior of the symmetry energy.

In Fig.~\ref{fig:EA}, we plot the energy per baryon $E/A$ for symmetric nuclear matter and 
pure neutron matter as a function of baryon density $n_b$ 
in the TM1e and TM1 models (upper panel) and in the IUFSU models (lower panel). 
It is seen that the behaviors of symmetric nuclear matter are the same between the models with 
different $L$, while significant differences are evident in pure neutron matter. 
In Fig.~\ref{fig:Esym}, we plot the symmetry energy $E_{\rm sym}$ as a function of $n_b$. 
The model with a smaller $L$ exhibits a slightly larger $E_{\rm sym}$ at low densities 
but a much smaller $E_{\rm sym}$ at high densities than the model with a larger $L$.
This discrepancy primarily arises from the different values of the symmetry energy slope
$L$ in these models. 
By making a detailed comparison between models with different $L$, we can conveniently explore 
how the symmetry energy and its density dependence influence the properties of nuclear pasta 
in stellar matter.

\begin{figure*}[htbp]
	\begin{center}
		\includegraphics[clip,width=13.0 cm]{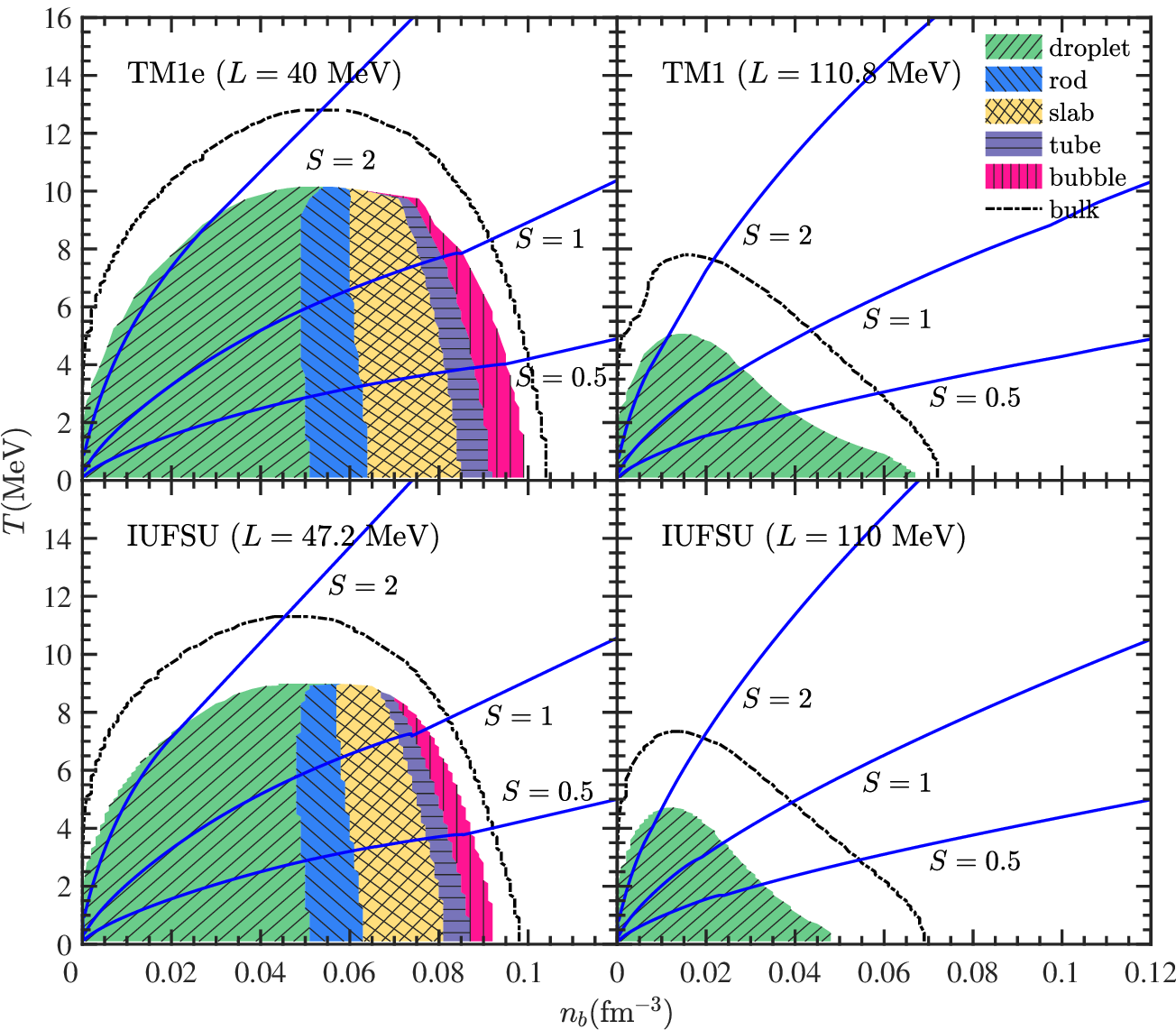}
		\caption{Phase diagrams in the $n_b$--$T$ plane obtained in the CLD method. 
			Different colors indicate the regions for different pasta shapes. 
			The black dot-dashed lines indicate the boundary of nonuniform matter 
            obtained from a bulk calculation without finite-size effects.
			The blue solid lines correspond to isentropic trajectories with entropy 
            per baryon $S=0.5,\, 1,\, 2$.}
		\label{fig:Tnb}
	\end{center}
\end{figure*}

\begin{figure*}[htbp]
	\begin{center}
		\includegraphics[clip,width=13.0 cm]{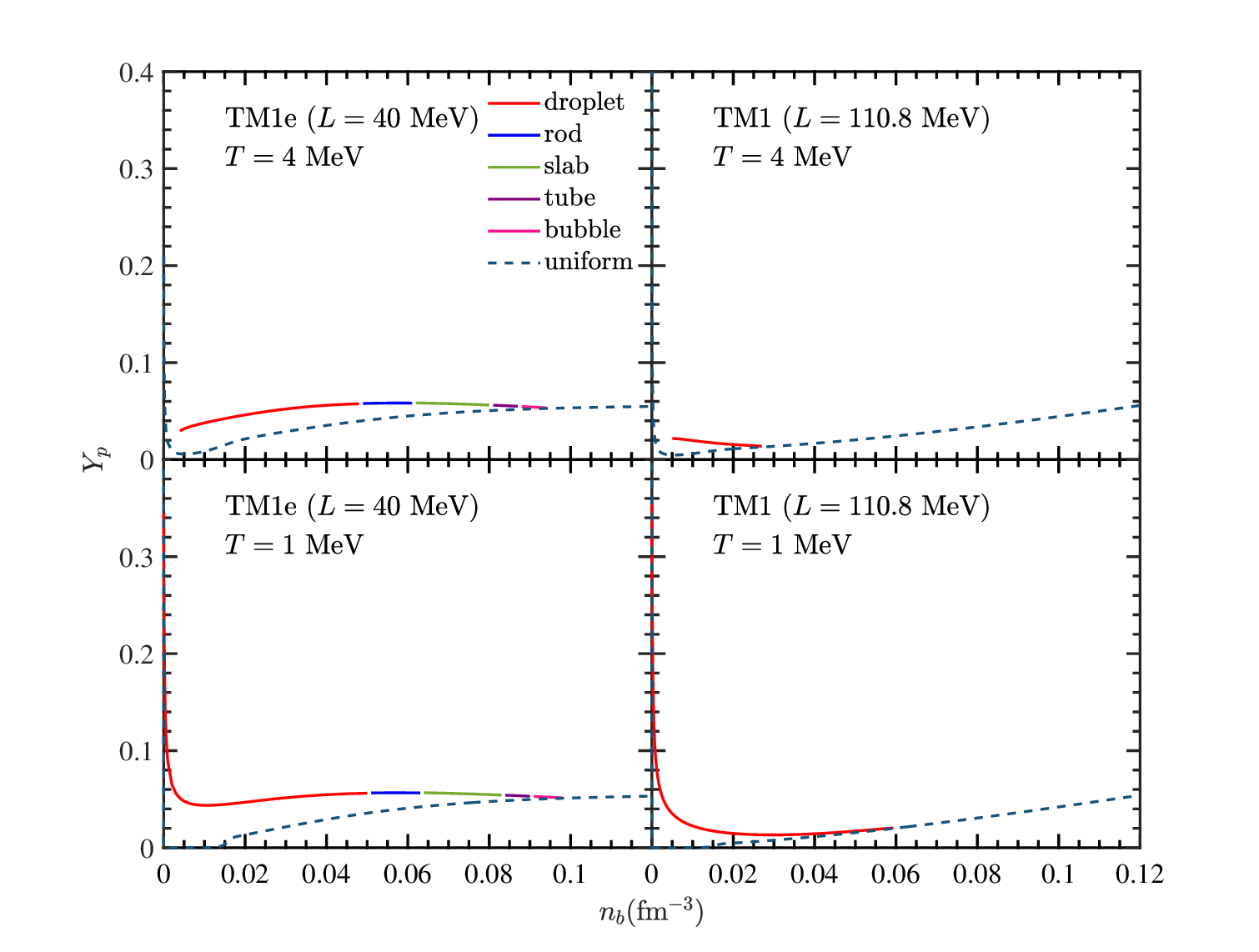}
		\caption{Average proton fraction $Y_p$ as a function of the baryon density $n_b$ 
			in matter under the conditions of charge neutrality and $\beta$ equilibrium. 
			The results for the TM1e and TM1 models are shown in the left and right panels, respectively. 
			The dashed lines indicate the proton fraction of uniform matter.}
		\label{fig:Yp}
	\end{center}
\end{figure*}

\begin{figure*}[htbp]
	\begin{center}
		\includegraphics[clip,width=13.0 cm]{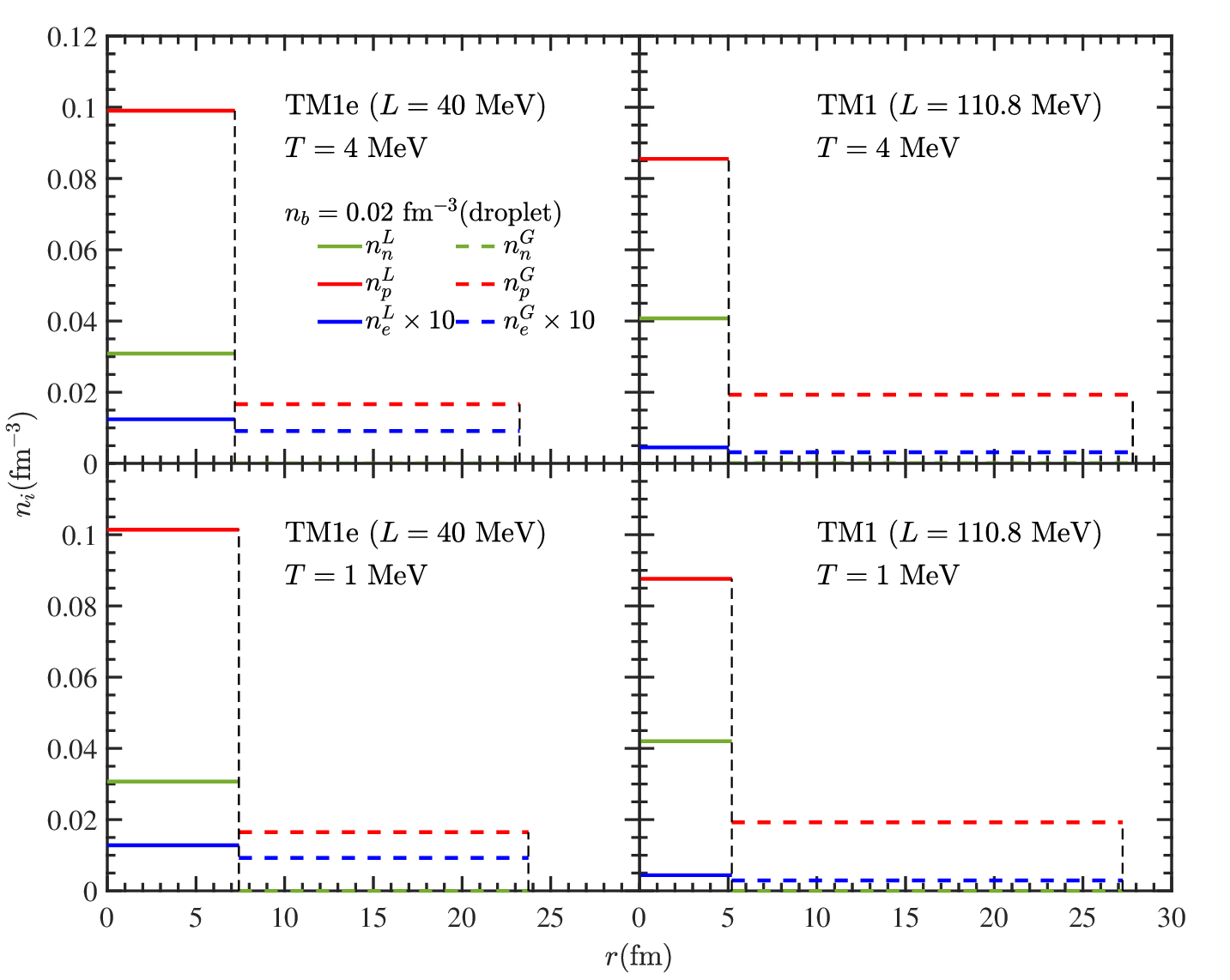}
		\caption{Density profiles of nucleons and electrons within the Wigner-Seitz cell 
			for a droplet configuration at $n_b=0.02 \, {\rm fm^{-3}}$. 
			The results obtained with the TM1e and TM1 models are shown in the left and right panels, respectively. 
			The cell boundary and the sharp interface between the liquid and gas phases are indicated 
			by the vertical dashed lines.}
		\label{fig:ni-r}
	\end{center}
\end{figure*}

\begin{figure*}[htbp]
	\begin{center}
		\includegraphics[clip,width=13.0 cm]{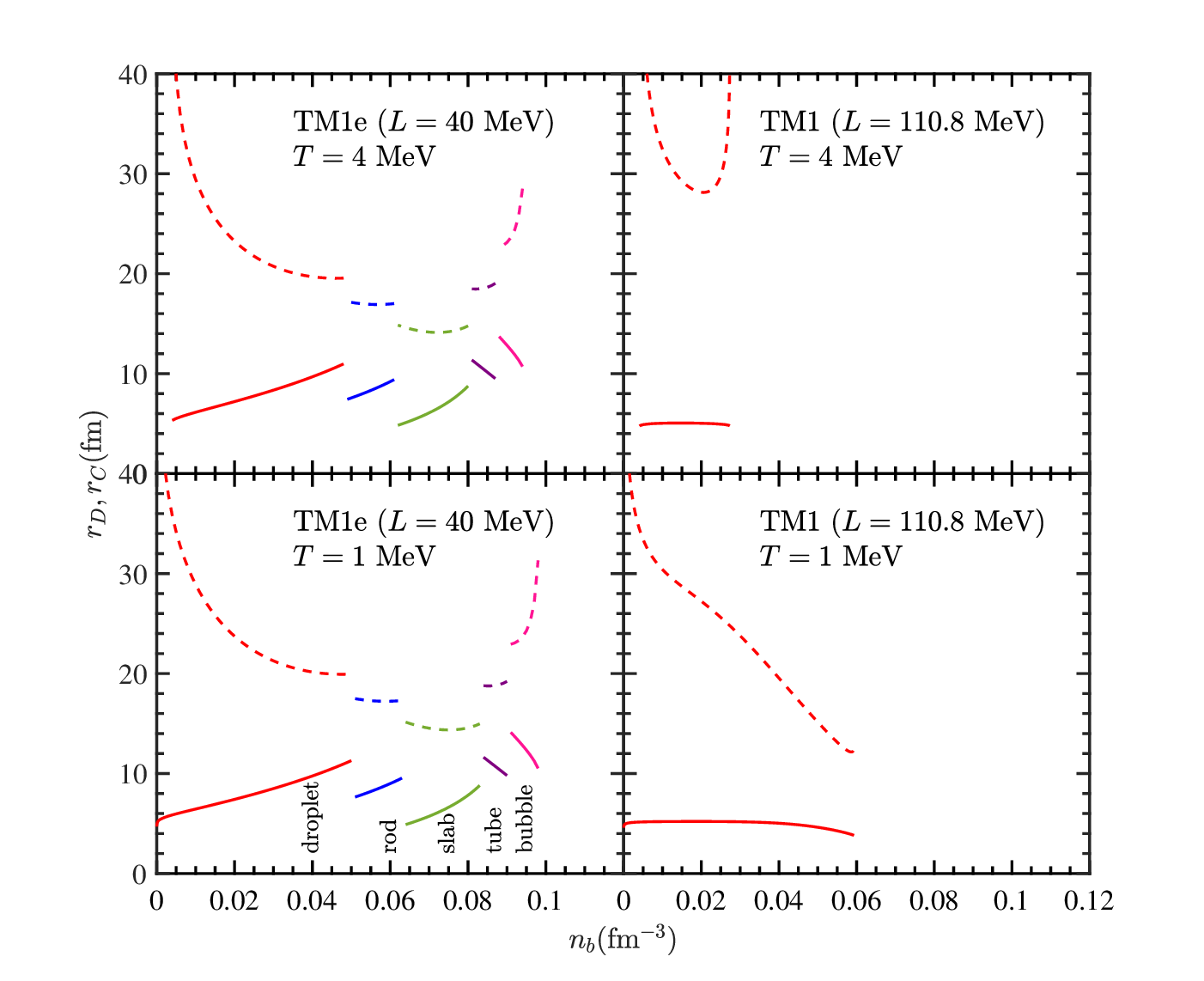}
		\caption{Size of the Wigner-Seitz cell $r_C$ (dashed lines) and that of the inner phase $r_D$ (solid lines) 
			as a function of the baryon density $n_b$. 
			The results obtained with the TM1e and TM1 models are shown in the left and right panels, respectively. }
		\label{fig:rdrc}
	\end{center}
\end{figure*}

\begin{figure*}[htbp]
	\begin{center}
		\includegraphics[clip,width=13.0 cm]{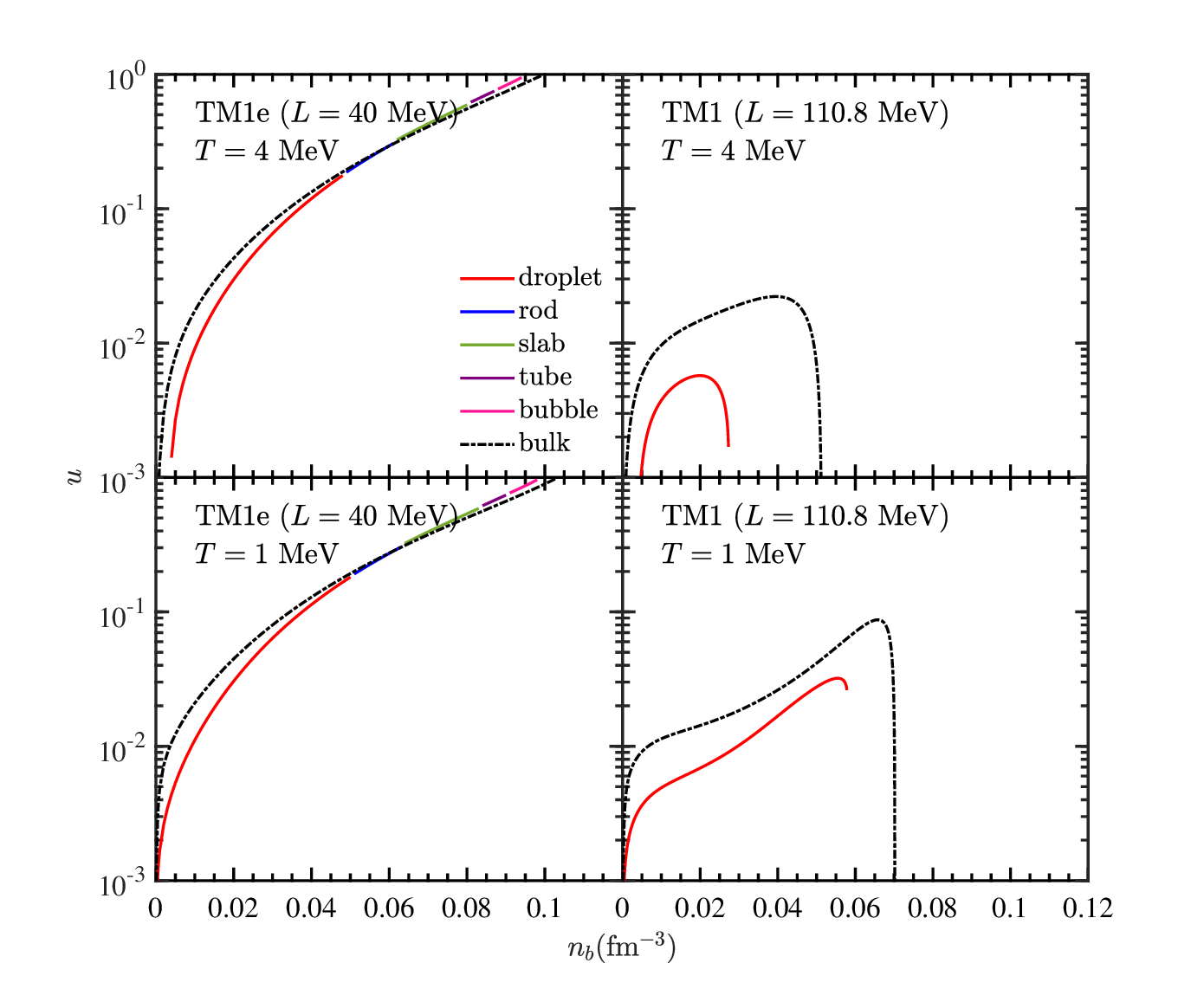}
		\caption{Volume fraction of the liquid phase $u$ as a function of the baryon density $n_b$.
			The black dot–dashed lines correspond to the bulk equilibrium results.}
		\label{fig:vol}
	\end{center}
\end{figure*}

\begin{figure*}[htbp]
	\begin{center}
		\includegraphics[clip,width=15.6 cm]{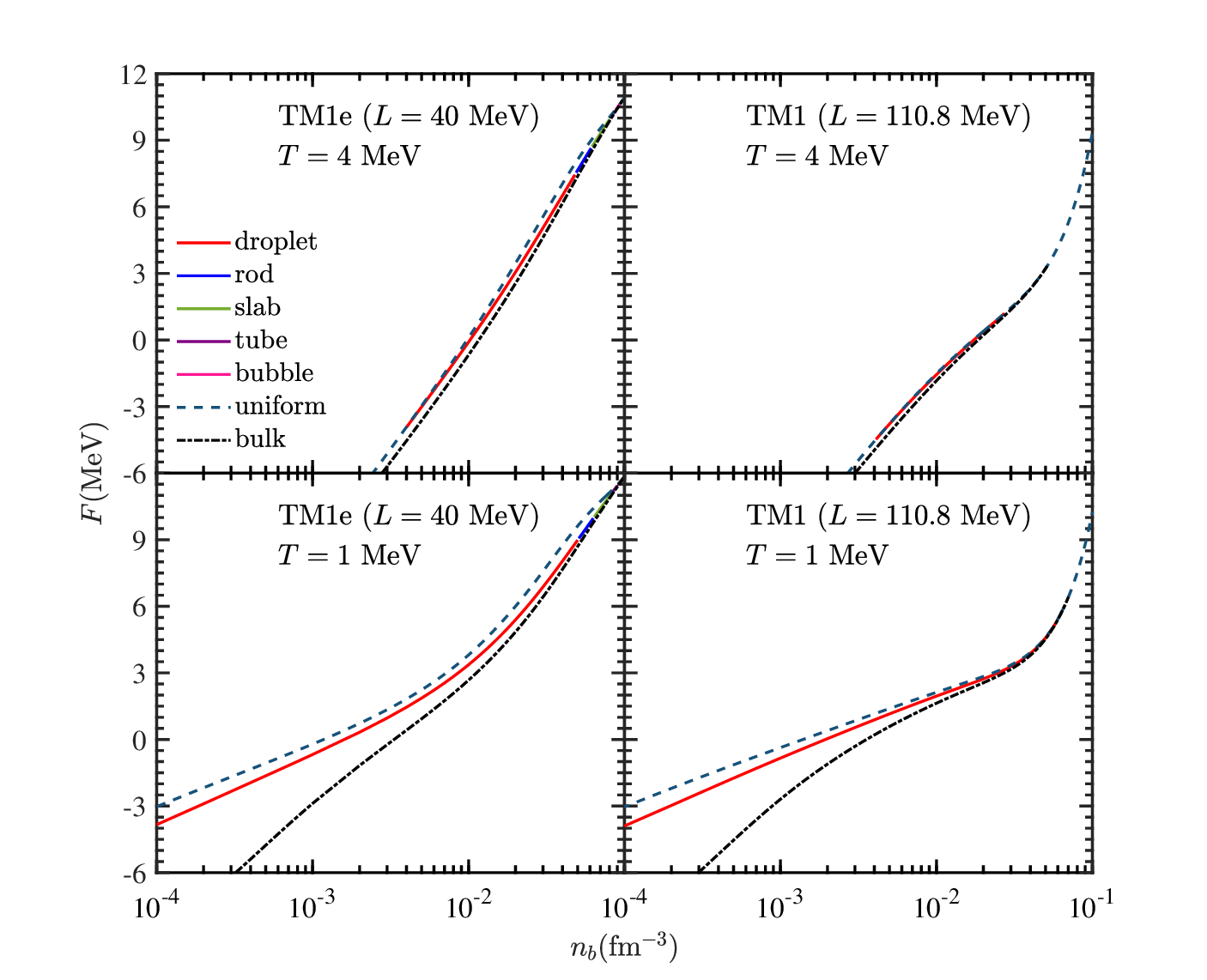}
		\caption{Free energy per baryon $F$ as a function of the baryon density $n_{b}$. 
			The dashed lines correspond to the results for uniform matter, while the dot-dashed lines
			indicate those from bulk equilibrium calculations.}
		\label{fig:FA}
	\end{center}
\end{figure*}

\begin{figure*}[htbp]
	\begin{center}
		\includegraphics[clip,width=15.6 cm]{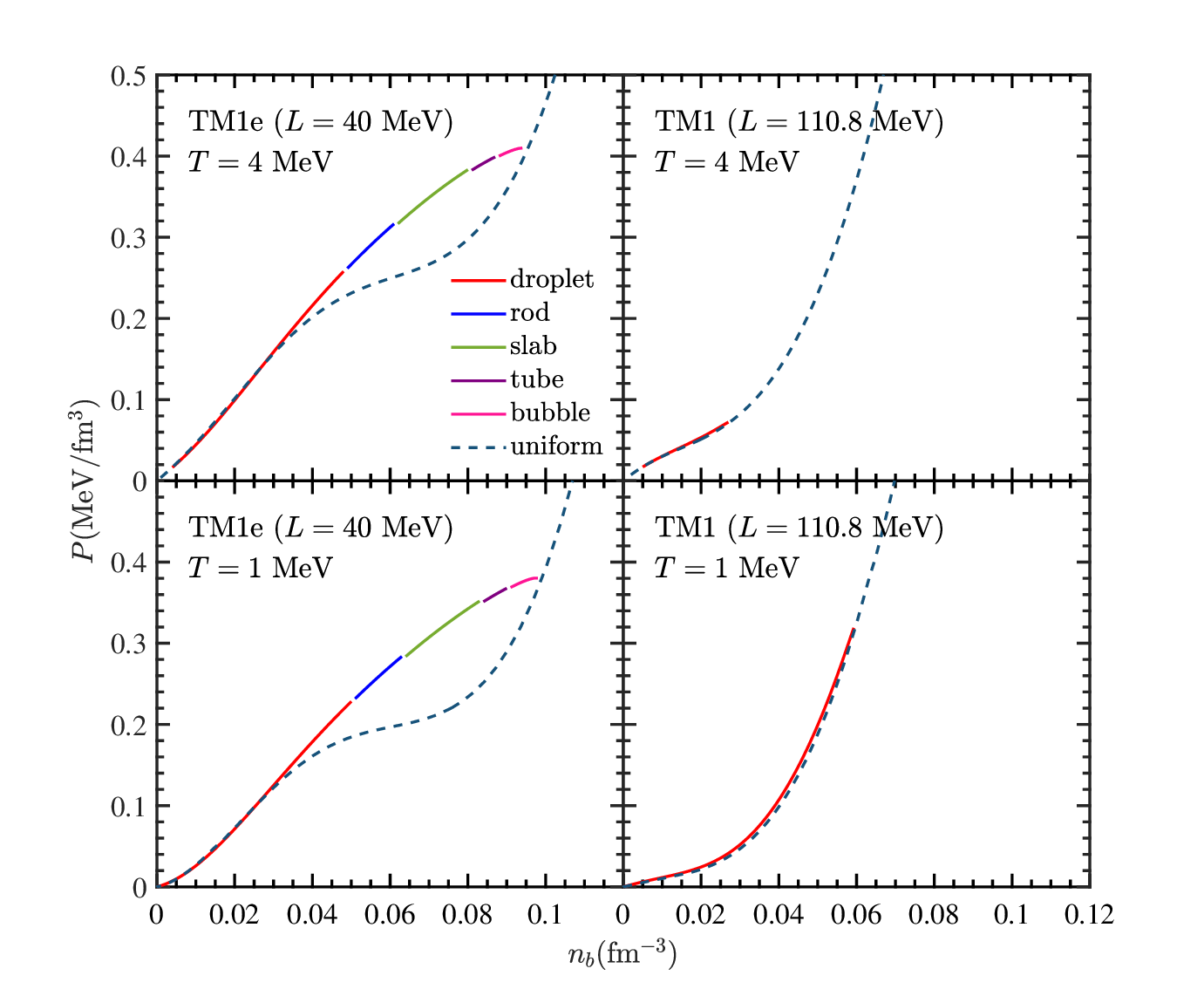}
		\caption{Pressure $P$ as a function of the baryon density $n_{b}$ obtained using the TM1e and TM1 models.
			The results of pasta phases (colored solid lines) are compared with those of uniform 
			matter (dashed lines).}
		\label{fig:pnb}
	\end{center}
\end{figure*}

\begin{figure*}[htbp]
	\begin{center}
		\includegraphics[clip,width=15.6 cm]{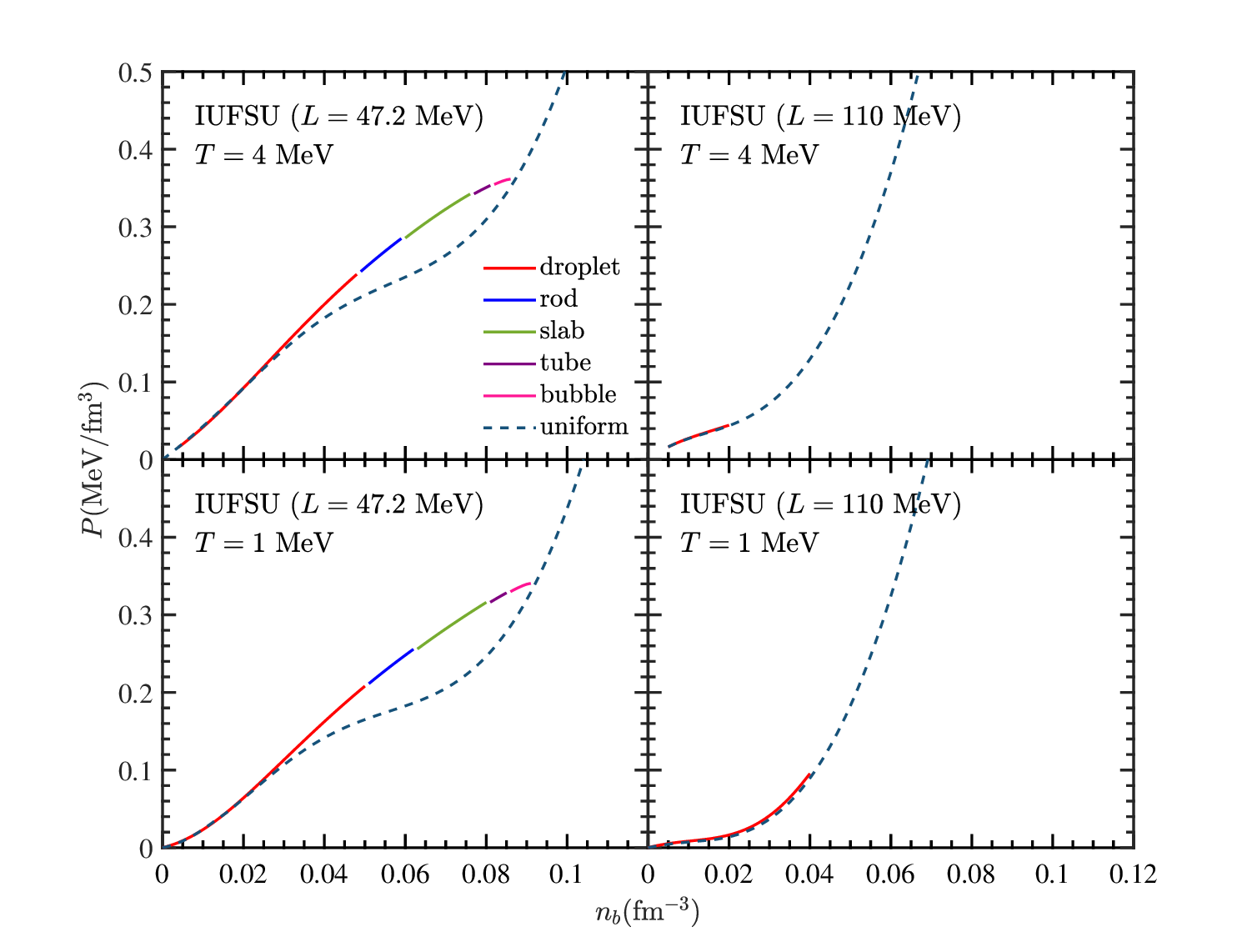}
		\caption{Same as Fig.~\ref{fig:pnb}, but for the IUFSU models. }
		\label{fig:pnb-IUFSU}
	\end{center}
\end{figure*}

\subsection{CLD method at finite temperature}
\label{sec:2.2}

We employ the CLD method to describe the pasta phases in hot neutron-star matter.
The Wigner-Seitz approximation is adopted for simplifying the calculations, 
in which the whole space is divided into charge-neutral cells with assumed geometric shapes.
In the CLD method, the matter within the Wigner-Seitz cell is assumed to consist of 
a dense liquid ($L$) and a dilute gas ($G$) phase, separated by a sharp interface.
Generally, the geometric shapes are likely to vary from droplet to rod, slab, tube, 
and bubble with increasing baryon density.

At a given temperature $T$ and average baryon density $n_{b}$, the equilibrium state of
neutron-star matter can be determined by minimizing
the total free energy density of the system among all configurations
considered~\cite{latt91,shen11,bao14b}.
The free energy density of the pasta phase is written as
\begin{eqnarray}
	\label{eq:free}
	f&=&uf^{L}(n_{p}^{L},n_{n}^{L},n_e^L)+\left(1-u\right)f^{G}(n_{p}^{G},n_{n}^{G},n_{\alpha}^{G},n_e^G) \notag\\
	&&+f_{\mathrm{surf}}\left(u,r_D,\tau\right)+f_{\mathrm{Coul}}(u,r_D,\delta n_c),
\end{eqnarray}
where $u$ denotes the volume fraction of the liquid phase,
$L$ and $G$ represent the liquid and gas phases, respectively.
The free energy densities of homogeneous matter in the liquid and gas phases, $f^{L}$ and $f^{G}$, 
account for the bulk contributions, which can be calculated using Eq.~\eqref{eq:fhp}.
It is noteworthy that $\alpha$ particles may exist as a representative light
nucleus in the dilute gas phase, whereas they are absent in the dense liquid phase.
Therefore, we include the contributions of $\alpha$ particles in the gas phase,
which are treated as noninteracting Boltzmann particles in the present calculation.
The last two terms of Eq.~\eqref{eq:free}, $f_{\mathrm{surf}}$ and $f_{\mathrm{Coul}}$, represent
the surface energy and Coulomb energy, respectively, which arise from the finite-size effects.
Here, $r_D$ denotes the size of the inner part inside the Wigner-Seitz cell with a geometric dimension $D$, 
while $\tau$ represents the surface tension. 
The quantity $\delta n_c=n_c^L-n_c^G$ denotes the charge density difference between 
the liquid and gas phases.

The finite-size effect plays a crucial role in the studies of pasta phases~\cite{bao16,ji20}. 
We include the surface and Coulomb terms in determining the equilibrium conditions for two-phase coexistence.
The surface and Coulomb energy densities are calculated from
\begin{eqnarray}
	&&f_{\mathrm{surf}}=\frac{D \tau u_{\mathrm{in}}}{r_D},\label{eq:surf}\\
	&&f_{\mathrm{Coul}}=\frac{e^{2}}{2}(\delta n_{c})^{2} r_D^2 
	u_{\mathrm{in}}\Phi(u_{\mathrm{in}})\label{eq:Coul},
\end{eqnarray}
with
\begin{eqnarray}
	\Phi(u_{\mathrm{in}})=
	\begin{cases}
		\dfrac{1}{D+2}\biggl(\dfrac{2-Du_{\mathrm{in}}^{1-2/D}}{D-2}+u_{\mathrm{in}}\biggr),&D=1,3,\\
		\dfrac{u_{\mathrm{in}}-1-\ln u_{\mathrm{in}}}{D+2},&D=2,
	\end{cases}
\end{eqnarray}
where $D=1,2,3$ is the geometric dimension of the cell.  
$u_{\mathrm{in}}$ represents the volume fraction of the inner part, i.e., 
$u_{\mathrm{in}}=u$ for droplet, rod, and slab configurations, and 
$u_{\mathrm{in}}=1-u$ for tube and bubble configurations. 
$e=\sqrt{4\pi /137}$ is the 
electromagnetic coupling constant. 

The surface tension $\tau $ is an essential quantity that can significantly affect 
the structure of pasta phases~\cite{avan10,ji20}. 
Generally, $\tau $ is achieved by using the Thomas-Fermi approximation 
for a one-dimensional nuclear system~\cite{avan10,bao14a}.
At finite temperature, both the surface energy and
surface entropy are included in the surface tension $\tau $.
In the present work, we employ a parameterized surface tension $\tau $,
which is a function of the temperature $T$ and the proton fraction in 
the dense liquid phase $Y_p^L$~\cite{schn17},
\begin{eqnarray}
	\tau(Y_p^L,T)=\tau_s h(Y_p^L,T)\frac{2\times2^\lambda+q}
	{\left(Y_p^L\right)^{-\lambda}+q+\left(1-Y_p^L\right)^{-\lambda}},
\end{eqnarray}
with $\tau_s \equiv \tau(0.5,0)$. The function $h(Y_p^L,T)$ is defined as
\begin{eqnarray}
	h\left(Y_p^L,T\right)=
	\begin{cases}
		\left\{1-\left[T/T_{c}(Y_p^L)\right]^{2}\right\}^{p}, & \hspace{0.02cm} T\le T_{c}(Y_p^L);
		\\
		\;0, & \hspace{0.02cm} T > T_{c}(Y_p^L).
	\end{cases}
\end{eqnarray}
The critical temperature $T_c$ depends on the proton fraction $Y_p^L$,
which is fitted using the function
\begin{equation}
\small
	T_c(Y_p^L)=T_{c0}[a_c+b_c\delta^2(Y_p^L)+c_c\delta^4(Y_p^L)+d_c\delta^6(Y_p^L)],
\end{equation}
where $T_{c0}\equiv T_c(Y_p^L=0.5)$ is the critical temperature for symmetric nuclear matter 
and $\delta(Y_p^L)=1-2Y_p^L$ is the neutron excess. 
The parameterized surface tension $\tau$ is determined by fitting numerical results of 
the Thomas-Fermi approximation for a one-dimensional nuclear system.
In Table~\ref{tab:tension}, we list the parameters in the surface tension formula 
for the TM1e, TM1, and IUFSU models.
To examine the reliability of the fitting procedure, we compare in Fig.~\ref{fig:suft} 
the parameterized results (red dashed lines) with the surface tensions obtained within 
the Thomas-Fermi approximation (blue solid lines).
It can be seen that the two sets of results are in good agreement.

\begin{table*}[htp]
	\caption{Parameters in the surface tension $\tau$ for the TM1e, TM1, and IUFSU models. }
	\label{tab:tension}
	\begin{center}
		\begin{tabular}{lcccccccccc}
			\hline\hline
			Model &$\tau_s$ & $T_{c0}$ &$a_c$&$b_c$&$c_c$&$d_c$&$q$&$\lambda$&$p$\\
			&(MeV/fm$^2$) & (MeV) & & & & & & & \\
			\hline
			TM1e ($L=40$)&1.071& 15.6 &1.004 &$-$0.273 &1.407 &$-$1.421&1.2546 & 1.8844&1.616 \\
			\hline
			TM1 ($L=110.8$)&1.071& 15.6 &1.076 &$-$1.119 &2.985 &$-$2.013&341.5 & 5.932&1.929 \\
			\hline
			IUFSU $(L=47.2)$&1.17& 14.5 &0.9903 &$-$0.1910 &0.6912 &$-$1.1695&1.5453 & 2.0802&1.4923\\
			\hline
			IUFSU $(L=110)$&1.17& 14.5 &1.0817 &$-$1.0154 &0.3146 &3.7947&480.2982 & 6.3908&1.9106\\
			\hline\hline
		\end{tabular}
	\end{center}
\end{table*}

For a given pasta shape at temperature $T$ and average baryon density $n_{b}$,
the equilibrium state can be determined by minimizing the total free energy density $f$ described 
in Eq.~\eqref{eq:free}, which is a function of the following variables: 
$n_{p}^{L}$, $n_{n}^{L}$, $n_{e}^{L}$, $n_{p}^{G}$, $n_{n}^{G}$, $n_{e}^{G}$, $n_{\alpha }^{G}$, $u$, 
and $r_D$. These variables satisfy the constraints of baryon number conservation and global charge 
neutrality:
\begin{eqnarray}
	u\left(n_p^L+n_n^L\right)+(1-u)\left(n_p^G+n_n^G+4n_\alpha^G\right) &=& n_b,\\
	u\left(n_p^L-n_e^L\right)+(1-u)\left(n_p^G+2n_\alpha^G-n_e^G\right) &=& 0.
	\label{eq:charge0}
\end{eqnarray}
We introduce the Lagrange multipliers $\mu_n$ and $\mu_e$ to enforce the constraints and minimize 
the function,
\begin{eqnarray}
	w &=& f-\mu _{n}\left[ u\left(n_p^L+n_n^L\right)+(1-u)\left(n_p^G+n_n^G+4n_\alpha^G\right) \right]
	\notag \\
	&+& \mu _{e}\left[ u\left(n_p^L-n_e^L\right)+(1-u)\left(n_p^G+2n_\alpha^G-n_e^G\right) \right],
\end{eqnarray}
which lead to the following equilibrium conditions:
\begin{eqnarray}
	\mu_n^G &=& \mu_n^L,  \label{eq:neu1}\\
	\mu_p^G &=& \mu_p^L+\frac{2f_{\rm Coul}}{u(1-u)\delta n_c}, \label{eq:LGp}\\
	\mu_e^G &=& \mu_e^L-\frac{2f_{\rm Coul}}{u(1-u)\delta n_c}, \label{eq:LGe}\\
	\mu_n^L &=& \mu_p^L+\mu_e^L, \label{eq:beta1}\\
	\mu_n^G &=& \mu_p^G+\mu_e^G, \label{eq:beta2}\\
	\mu_\alpha^G &=& 2\mu_p^G+2\mu_n^G, \label{eq:alpha}\\
	P^G &=& P^L+\frac{2f_{\rm Coul}}{\delta n_c}\left(\frac{n_c^L}{u}+\frac{n_c^G}{1-u}\right) \notag\\
	& & \mp \left[\frac{f_{\rm surf}}{u_{\rm in}}+\frac{f_{\rm Coul}}{u_{\rm in}}
	\left(1+u_{\rm in}\frac{\Phi^\prime}{\Phi}\right)\right]. \label{eq:pre}
\end{eqnarray}
The sign in the last equation is \textquotedblleft $-$\textquotedblright\ for droplet, rod,
and slab configurations, while it is \textquotedblleft $+$\textquotedblright\ for tube and 
bubble configurations. It is shown in Eqs.~\eqref{eq:beta1} and~\eqref{eq:beta2} that 
local $\beta$ equilibrium can be reached in both liquid and gas phases.
The terms involving $f_{\rm Coul}$ and $f_{\rm surf}$ arise from finite-size effects
and reflect the differences between the CLD method and the Gibbs phase equilibrium conditions.

We can obtain the equilibrium condition $f_{\rm surf}=2f_{\rm Coul}$ by minimizing $w$ with respect 
to $r_D$. Consequently, the size of the inner part is given by
\begin{eqnarray}
	\label{eq:rD}
	r_D=\left[\frac{D \tau}{e^2(\delta n_c)^2\Phi}\right]^{1/3},
\end{eqnarray}
whereas the size of the Wigner--Seitz cell is obtained from
\begin{eqnarray}
	\label{eq:rC}
	r_C = u_{\rm{in}}^{-1/D} r_D.
\end{eqnarray}

In practice, we solve the equilibrium equations at a given temperature $T$ and average baryon 
density $n_{b}$ for all pasta configurations, and then identify the state with the lowest 
free energy density as the thermodynamically stable phase.
When the free energy density of pasta phases is higher than that of uniform matter,
the system prefers to have a homogeneous distribution.
In the pasta phases, the pressure and chemical potentials of the system
may be different from those in the liquid and gas phases due to finite-size effects.
Therefore, we compute these quantities by the thermodynamic relations.

\begin{figure*}[htbp]
	\centering
	\includegraphics[clip,width=15.6 cm]{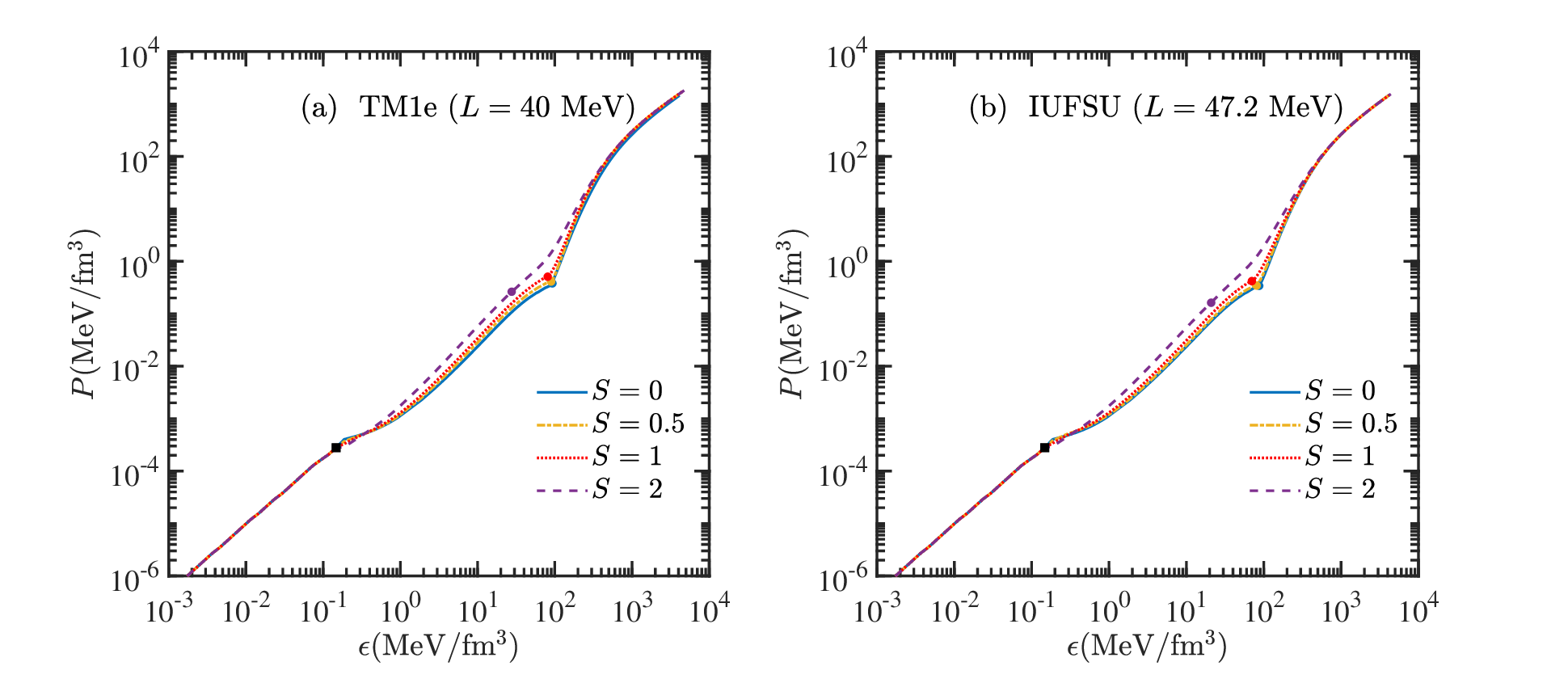}
	\caption{Pressure $P$ as a function of the energy density $\epsilon$ for several values 
		of entropy per baryon $S$ obtained using (a) the TM1e ($L=40$ MeV) model
		and (b) the IUFSU ($L=47.2$ MeV) model. 
		The cold BPS EOS is adopted for the outer crust, and the matching 
		point with the inner crust is marked by the black square. 
		The crust-core transition is indicated by the colored circles.}
	\label{fig:pelog}
\end{figure*}

\begin{figure*}[htbp]
	\centering
	\includegraphics[clip,width=15.6 cm]{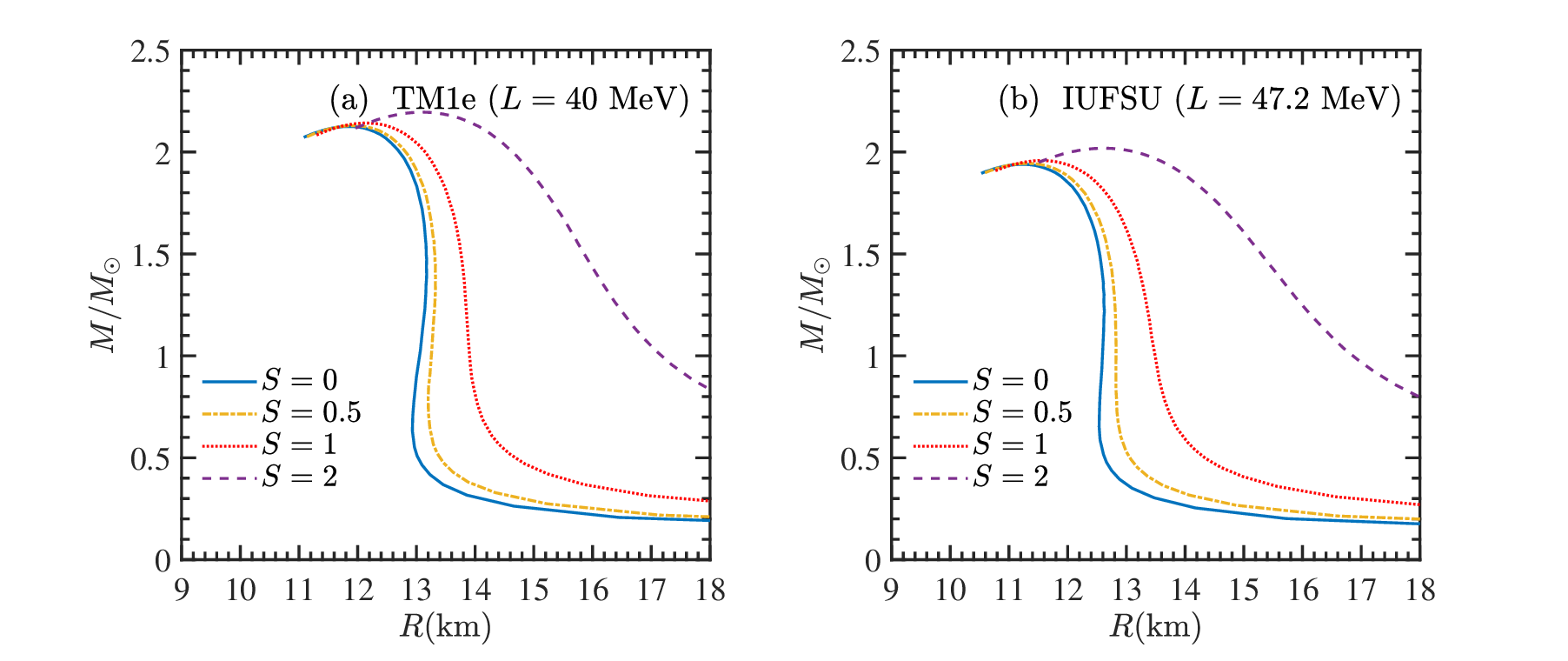}
	\caption{Mass-radius relations of PNSs obtained using the isentropic EOS 
		shown in Fig.~\ref{fig:pelog}.}
	\label{fig:RvsM}
\end{figure*}

\begin{figure*}[htbp]
	\centering
	\includegraphics[clip,width=15.6 cm]{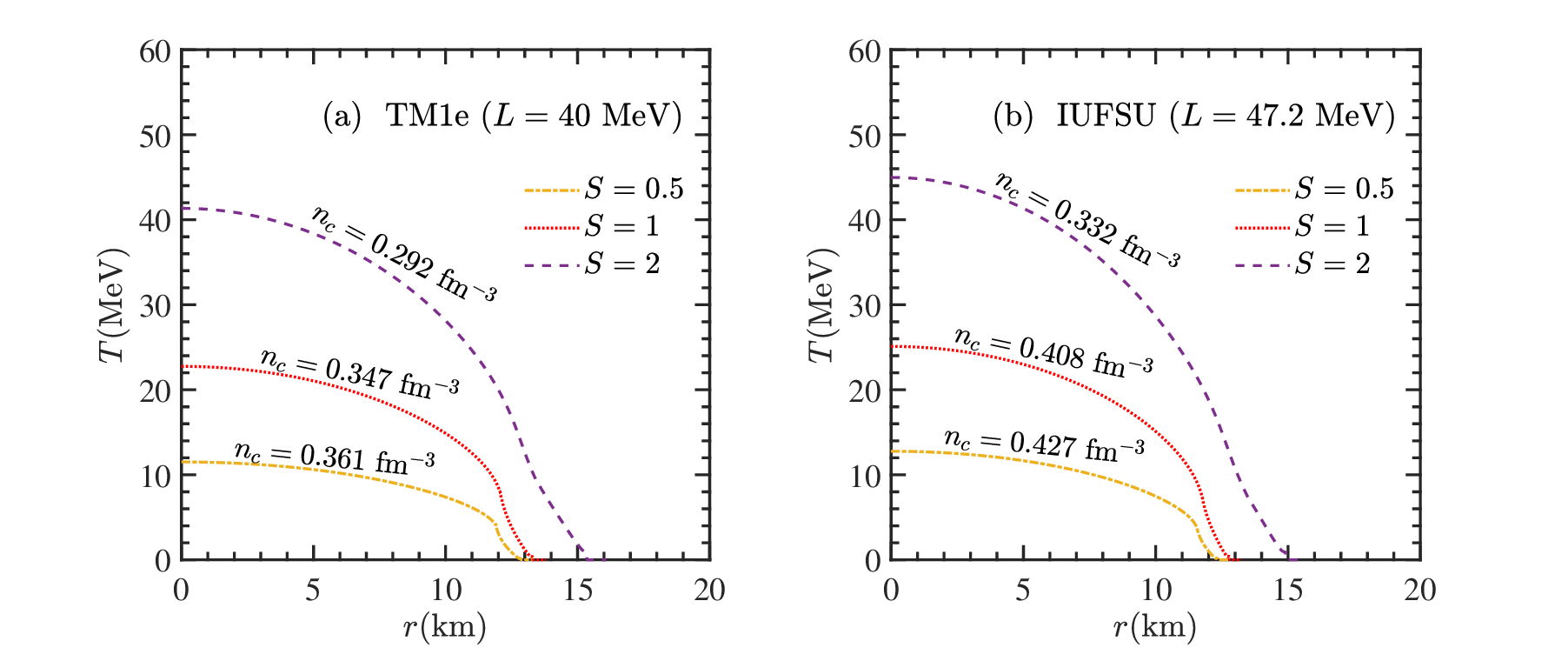}
	\caption{Temperature profiles of a canonical $1.4\ \Msun$ PNS 
		obtained with the constant entropy approach for $S=0.5,\, 1,\, 2$. 
		The corresponding central density $n_c$ increases as $S$ decreases. }
	\label{fig:TR}
\end{figure*}

\begin{figure*}[htbp]
	\centering
	\includegraphics[clip,width=15.6 cm]{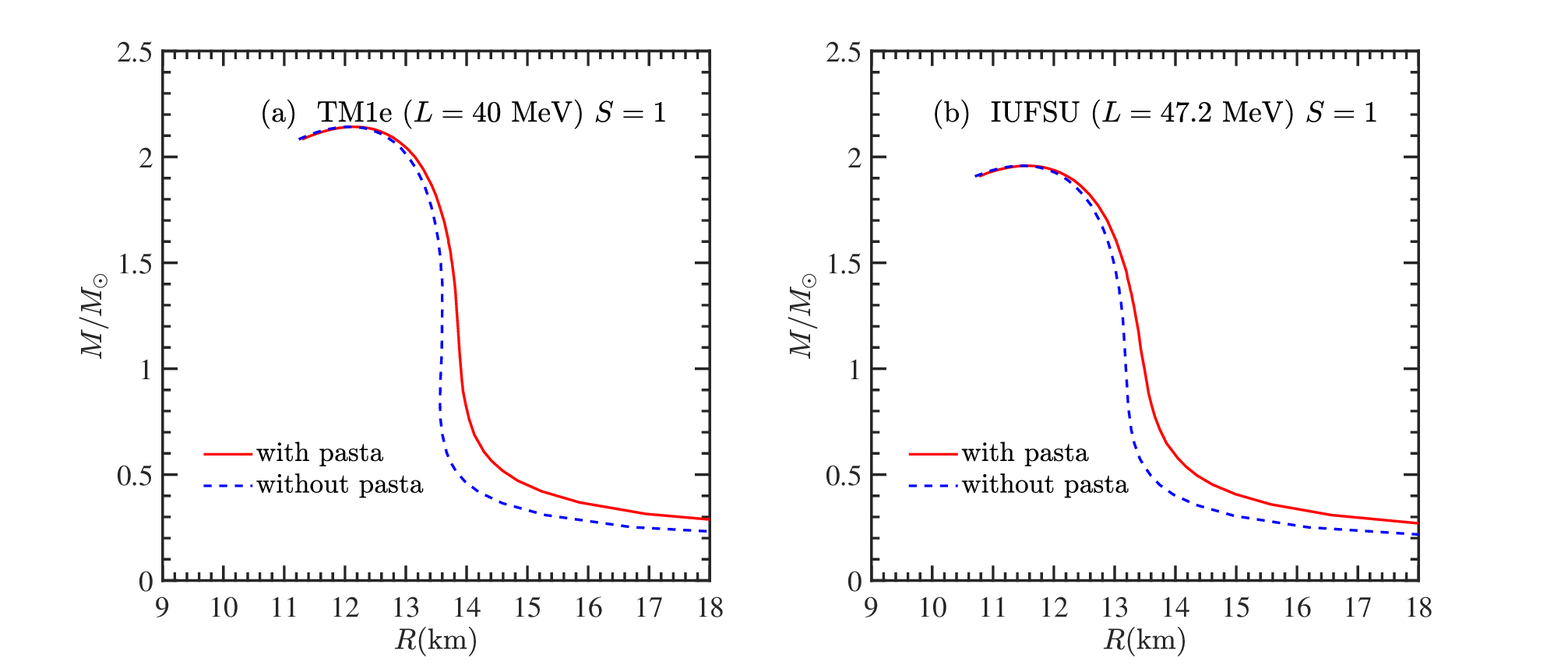}
	\caption{Mass-radius relations obtained using the isentropic EOS for $S=1$,
		including pasta phases in the inner crust (red solid line). 
		The results without pasta phases (blue dashed line) are shown for comparison. }
	\label{fig:MRwo1}
\end{figure*}

\begin{figure*}[htbp]
	\centering
	\includegraphics[clip,width=15.6 cm]{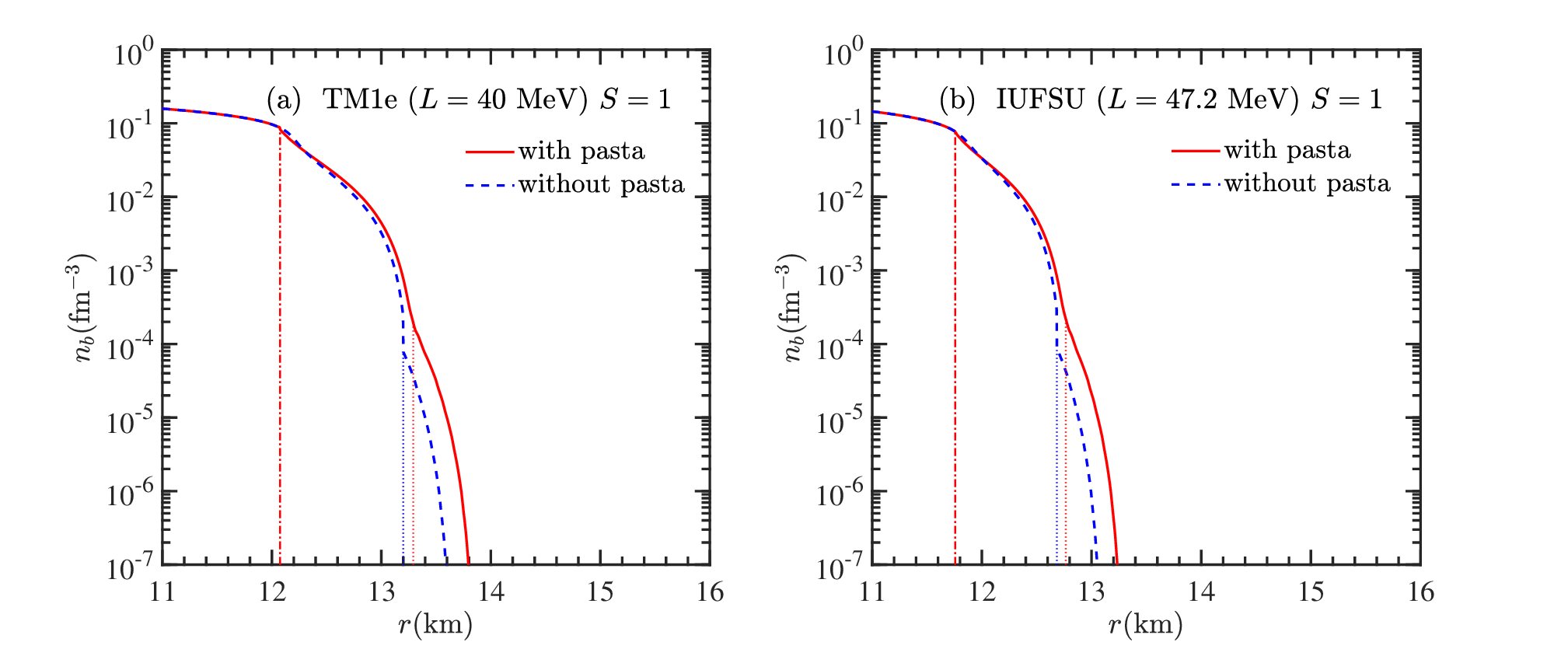}
	\caption{Density profiles in the surface region of a canonical $1.4\Msun$ PNS for $S=1$. 
		The red vertical dot-dashed line indicates the boundary between the uniform core 
		and the inner crust, whereas the red dotted line marks the boundary between the inner 
		crust and the outer crust. In the case without pasta phases, the uniform core connects 
		directly to the outer crust indicated by the blue vertical dotted line.}
	\label{fig:nb-ro1}
\end{figure*}

\section{Results and discussion}
\label{sec:3}

\subsection{Nuclear pasta in hot neutron-star matter}
\label{sec:3.1}

We first investigate the pasta phases appearing in nuclear matter under $\beta$ equilibrium 
at finite temperature, which is considered to occur in the crust of a PNS.
To describe these pasta phases, we employ the CLD method, in which a dense nuclear liquid coexists
with a dilute gas across a sharp interface.
Regarding the nuclear interaction, 
both the TM1e model ($L=40$ MeV) and the IUFSU model ($L=47.2$ MeV) are compatible 
with nuclear experimental data and recent astrophysical observations of neutron stars. 
To examine the impact of the symmetry energy and its slope, 
we perform a comparison between a pair of RMF models with different values of $L$. 
It should be noted that the two models in each pair differ only in their isovector parts, 
while their isoscalar properties remain identical.

At a given temperature $T$ and average baryon density $n_b$, we perform calculations for all 
configurations including both uniform matter and nonuniform pasta phases, and then identify 
the one with the lowest free energy density as the thermodynamically stable phase.
In Fig.~\ref{fig:Tnb}, we show the phase diagrams in the $n_b$--$T$ plane obtained using  
the two pairs of models considered in this work.
The blue solid lines correspond to isentropic trajectories with entropy per baryon $S=0.5,\, 1,\, 2$.
Inside the nonuniform matter region, the values of $S$ are taken to be the cell-averaged entropy per baryon.
One can see that the models with smaller values of $L$, namely TM1e ($L=40$ MeV) and IUFSU ($L=47.2$ MeV),
exhibit various pasta phases at low temperatures, 
whereas the models with larger values of $L$, namely TM1 ($L=110.8$ MeV) and IUFSU ($L=110$ MeV), 
exhibit only the droplet phase.
This difference arises because the symmetry energy slope $L$ can significantly influence 
the emergence of pasta phases.
A similar correlation between $L$ and pasta phases has also been observed in the studies of
the neutron-star inner crust~\cite{bao15}.
In the left panels of Fig.~\ref{fig:Tnb}, we observe that various pasta configurations emerge 
sequentially with increasing density, while the transitions between different shapes are only 
weakly dependent on temperature $T$. As the temperature increases, the density range of
nonuniform matter shrinks. Eventually, when the temperature reaches the critical value $T_c$,
the nonuniform matter phase disappears completely.
In contrast, in the right panels, the critical value $T_c$ and nonuniform matter region 
are obviously smaller than those in the left panels.
This difference can be attributed to the different symmetry energy slopes in these models, 
which significantly influence the properties of $\beta$-equilibrium matter. 
The black dot-dashed lines indicate the boundary obtained from bulk calculations, 
where the finite-size effects such as surface and Coulomb contributions are neglected. 
Notably, the finite-size effects can significantly reduce the nonuniform matter region.
One can see that the two pairs of models shown in the upper and lower panels exhibit 
similar features, although quantitative differences remain.

In Fig.~\ref{fig:Yp}, we show the average proton fraction $Y_p$ as a function of the baryon 
density $n_b$ in matter under the conditions of charge neutrality and $\beta$ equilibrium. 
The results of the TM1e model (left panels) at $T=1$ and $4$ MeV are compared with those in the 
TM1 model (right panels). Because of different symmetry energy behaviors shown in Fig.~\ref{fig:Esym},
$Y_p$ in the TM1e model is significantly larger than that in the TM1 model.
Generally, a larger symmetry energy $E_{\rm sym}$ favors more symmetric nuclear matter, 
yielding a larger proton fraction $Y_p$ at the same density.
In nonuniform matter, the proton fraction in the dense liquid phase ($Y_p^{L}$) is much larger
than that in the gas phase ($Y_p^{G}$). Consequently, the average proton fraction $Y_p$ of 
the pasta phases (colored solid lines) is higher than that of uniform matter (dashed lines).
Furthermore, the results exhibit only a weak dependence on the temperature $T$. 
At $T=1$ MeV, $Y_p$ of nonuniform matter rises rapidly with decreasing density 
for $n_b < 0.05\, {\rm fm^{-3}}$. This trend is not observed at $T=4$ MeV,
since the onset of nonuniform matter occurs at a density of about $0.05$ fm$^{-3}$ in this case.

In Fig.~\ref{fig:ni-r}, we plot the particle density profiles within the Wigner-Seitz cell
for a droplet configuration at $n_b=0.02\, {\rm fm^{-3}}$ to analyze the equilibrium 
between the liquid and gas phases.
Particle densities in the dense liquid phase are shown by colored solid lines,
while those in the dilute gas phase are represented by colored dashed lines. 
The cell boundary and the sharp interface between the liquid and gas phases are indicated 
by the vertical dashed lines.
The dense liquid phase has a higher proton density ($n_p^{L}$) than the electron density ($n_e^{L}$),
so it is positively charged. In contrast, the dilute gas phase is negatively charged.
Furthermore, the electron density in the liquid phase ($n_e^{L}$) slightly exceeds 
that in the gas phase ($n_e^{G}$). This arises from the charge screening effect, which tends 
to reduce the net charge density in each phase to minimize the Coulomb energy.
The proton density in the gas phase ($n_p^{G}$) is significantly lower, which implies 
its proton fraction ($Y_p^{G}$) is much smaller than that in the liquid phase ($Y_p^{L}$).
Compared with the results in the TM1 model (right panels), the TM1e model yields
a larger nuclear radius ($r_D$), a higher neutron density ($n_n^{L}$), and a lower proton density ($n_p^{L}$).
These trends are attributed to the different symmetry-energy behaviors shown in Fig.~\ref{fig:Esym}.
A comparison of the upper and lower panels shows that the dependence on temperature $T$ is weak.

In Fig.~\ref{fig:rdrc}, we show the size of the Wigner--Seitz cell $r_C$ (dashed lines) 
and that of the inner phase $r_D$ (solid lines) as a function of the baryon density $n_b$.
Compared with various pasta phases obtained in the TM1e model (left panels), 
the TM1 model predicts only droplet configurations (right panels).
In the left panels, we see that $r_D$ in the droplet, rod, and slab phases 
increases with increasing $n_b$, whereas it decreases in the tube and bubble phases. 
The cell size $r_C$ exhibits the opposite trend.
This behavior is consistent with an increasing volume fraction of the liquid phase obtained
in the TM1e model (see the left panels of Fig.~\ref{fig:vol}).
In the right panels, the results for $T=4$ MeV are significantly different from those for $T=1$ MeV.
The density range of the droplet configuration for $T=4$ MeV is appreciably smaller than that for $T=1$ MeV.
This is consistent with the phase diagrams shown in Fig.~\ref{fig:Tnb}.
We display in Fig.~\ref{fig:vol} the volume fraction of the liquid phase $u$ as a function of 
the baryon density $n_b$. The black dot-dashed lines correspond to the bulk equilibrium results,
where finite-size effects are neglected. These effects are more pronounced in the 
TM1 model (right panels) than in the TM1e model (left panels).
It is shown that, in the TM1e model, $u$ monotonically increases with increasing $n_b$.
In contrast, $u$ in the TM1 model exhibits a nonmonotonic feature, where $u$ reaches a maximum
and then declines, implying that the nonuniform matter leaves the coexistence region 
in the dilute gas phase. This feature, called retrograde condensation~\cite{mull95,rogg18}, 
becomes more pronounced at higher temperatures.

In Fig.~\ref{fig:FA}, we display the free energy per baryon, $F$, as a function of the baryon
density $n_b$. The results shown by dashed lines for uniform matter are significantly higher 
than those represented by solid lines for nonuniform pasta phases, implying that the formation 
of pasta structures lowers the free energy at subsaturation densities.
On the other hand, the results of bulk equilibrium calculations shown by dot-dashed lines 
are lower than those of pasta phases, because the positive contributions from the surface and 
Coulomb terms are neglected in the bulk equilibrium approach.
The differences among these results are more pronounced at $T=1$ MeV (lower panels) than 
at $T=4$ MeV (upper panels), since the finite-size effects play a more important role at lower temperatures.
A comparison of the results in the TM1e model (left panels) with those in the TM1 model (right panels) 
shows that their trends are very similar, although quantitative differences remain.

In Figs.~\ref{fig:pnb} and~\ref{fig:pnb-IUFSU}, we present the pressure $P$ as a function of 
the baryon density $n_b$ obtained using the two pairs of models, respectively.
Compared with the TM1 model shown in the right panels of Fig.~\ref{fig:pnb}, 
the TM1e model shown in the left panels exhibits larger discrepancies
between uniform matter and nonuniform pasta phases. 
This is because the TM1e model with a smaller $L$ can yield a wider pasta  
region (see Fig.~\ref{fig:Tnb}) and a larger volume fraction $u$ (see Fig.~\ref{fig:vol}),
implying that pasta phases have a more significant impact on the results of the model with a smaller $L$.
Similar features are also observed for the IUFSU models in Fig.~\ref{fig:pnb-IUFSU}.

\subsection{Existence of nuclear pasta in a PNS}
\label{sec:3.2}

We investigate the influence of pasta phases on the macroscopic properties of PNSs,
where a constant entropy approach is adopted.
We consider the neutrino-free stage of PNSs with fixed entropy per baryon $S$.
In Fig.~\ref{fig:pelog}, we show the isentropic EOS for $S=0.5,\, 1,\, 2$, along with the $S=0$ curve
corresponding to the zero temperature case for comparison. 
As indicated in the phase diagram (Fig.~\ref{fig:Tnb}), the temperature $T$ along the 
isentropic trajectories (blue solid lines) decreases with decreasing baryon density $n_b$,
implying a decrease in temperature from the interior toward the surface of PNSs.
Accordingly, the system enters the nonuniform matter region at lower densities, 
corresponding to the inner crust of PNSs. 
For the outer crust, we adopt the zero-temperature BPS EOS, since the temperature 
in this regime is negligible.
In Fig.~\ref{fig:pelog}, the black square indicates the junction between the BPS EOS and the inner 
crust segments, while the colored circle marks the crust-core transition. It is observed that, 
for a given energy density, the pressure $P$ decreases monotonically with decreasing entropy $S$.

Using the isentropic EOS shown in Fig.~\ref{fig:pelog}, we solve the well-known 
Tolman-Oppenheimer-Volkoff (TOV) equation to obtain the properties of PNSs.
In Fig.~\ref{fig:RvsM}, the mass-radius relations of PNSs are shown for 
entropy per baryon $S=0,\, 0.5,\, 1,\, 2$. It is found that, as $S$ increases, both the 
maximum mass and the corresponding radius increase, and the mass-radius curve becomes flatter.
This implies that PNSs contract with the decrease of $S$ during the cooling process. 
In particular, the radii of lower-mass stars exhibit stronger sensitivity to $S$.
In Fig.~\ref{fig:TR}, we show the temperature profiles of a canonical $1.4\ \Msun$ PNS,
calculated using the constant entropy approach with $S=0.5,\, 1,\, 2$. 
It is found that the central density $n_c$ increases as $S$ decreases. 
For a fixed value of $S$, the temperature decreases with increasing radial coordinate $r$. 
Overall, a higher $S$ corresponds to higher temperatures in the stellar interior.

To investigate the influence of pasta phases on the macroscopic properties of PNSs, 
we plot in Fig.~\ref{fig:MRwo1} the mass-radius relation obtained using the isentropic EOS 
for $S=1$ with pasta phases appearing in the inner crust, and compare it with that obtained using 
the EOS of uniform matter without pasta phases.
It is observed that the radii with pasta phases are larger than those without pasta phases,
and this difference is more pronounced for lower-mass stars. This can be attributed to higher 
pressures in the presence of nonuniform pasta phases, compared to the uniform matter 
case (see the left panels of Figs.~\ref{fig:pnb} and~\ref{fig:pnb-IUFSU}).
In Fig.~\ref{fig:nb-ro1}, we show the density profiles in the surface region of a 
canonical $1.4 \Msun$ PNS. 
In the case with pasta phases, the red vertical dot-dashed lines indicates
the boundary between the uniform core and the inner crust, whereas the boundary between the inner
crust and the outer crust is indicated by the red dotted lines. 
Therefore, pasta phases exist in the inner crust, which has a thickness of about $1.0$--$1.2$ km.
In the case without pasta phases, we use the uniform EOS up to the outer crust, as indicated
by the blue vertical dotted lines.
In both cases, the outer crust exhibits similar behavior, but the presence of pasta phases
in the inner crust pushes the outer crust to larger radii. Furthermore, the pasta phases 
may change the thermal conductivity and influence the thermal evolution of the star~\cite{ghos24}.

\section{Conclusions}
\label{sec:4}

In this work, we have investigated the properties of nuclear pasta appearing in hot 
neutron-star matter. We employed the CLD method to describe the pasta phases 
with various geometric shapes. In the CLD method, the Wigner-Seitz approximation was 
employed to simplify the calculations, where the whole space was divided into 
equivalent cells of specific geometric symmetry. The matter in each cell consists 
of a dense liquid phase and a dilute gas phase separated by a sharp interface.
We employed a parametrized surface tension $\tau$ dependent on temperature and 
isospin asymmetry, with the parameters adjusted to match the results from the 
Thomas-Fermi approximation for a one-dimensional nuclear system.
We have derived the equilibrium conditions for two coexisting phases by minimizing the total 
free energy, which includes surface and Coulomb contributions. These conditions, incorporating 
finite-size effects, differ from the Gibbs conditions for phase equilibrium.
When finite-size effects are neglected, the CLD method reduces to a bulk matter calculation
that satisfies the Gibbs equilibrium conditions.

Regarding the nuclear interaction, we have employed two pairs of RMF models to examine 
the effects of the symmetry energy and its slope on pasta phases in PNSs.
The first pair consists of the original TM1 model with $L=110.8$ MeV and its extended 
version TM1e with $L=40$ MeV. The second pair consists of the original IUFSU model  
with $L=47.2$ MeV and a modified IUFSU model with $L=110$ MeV.
We have found that the models with smaller values of $L$ result in various pasta phases 
at low temperatures, while the models with larger values of $L$ yield only the droplet
configuration up to the crust-core transition density.
In addition, the properties of nonuniform neutron-star matter exhibit significant 
differences between models with different slope parameters.
It should be noted that the two models in each pair differ only in their isovector parts, 
which are closely related to the symmetry energy and its density dependence, 
while their isoscalar properties remain identical.
Therefore, the comparison between the two models in each pair allows us to isolate the influence 
of the symmetry energy and its slope, without interference from differences 
in the isoscalar properties.

We have explored the occurrence and influence of nuclear pasta in a PNS, 
employing a constant entropy approach. 
Using the isentropic EOS including pasta phases, 
we have solved the TOV equation to investigate the properties of PNSs. 
Nuclear pasta phases can appear in the inner crust with a thickness of 
about $1.0$--$1.2$ km. Compared with the results without pasta phases,
PNS radii including pasta phases are slightly larger, and this difference is more pronounced 
for lower-mass stars.
In the present calculations, we considered only neutrino-free matter in $\beta$ equilibrium,
corresponding to the later stages of the PNS cooling process.
Pasta phases in hot neutron-star matter with trapped neutrinos 
will be investigated in future work.

\section{Acknowledgments}

This work was supported by the National Natural Science Foundation of
China under Grants No. 12475149 and No. 12175109, 
and by the Guangdong Basic and Applied Basic Research Foundation (Grant  No. 2024A1515010911).

\bibliographystyle{apsrev4-1}
\bibliography{refs-zhou}

\begin{thebibliography}{60}%
\makeatletter
\providecommand \@ifxundefined [1]{%
 \@ifx{#1\undefined}
}%
\providecommand \@ifnum [1]{%
 \ifnum #1\expandafter \@firstoftwo
 \else \expandafter \@secondoftwo
 \fi
}%
\providecommand \@ifx [1]{%
 \ifx #1\expandafter \@firstoftwo
 \else \expandafter \@secondoftwo
 \fi
}%
\providecommand \natexlab [1]{#1}%
\providecommand \enquote  [1]{``#1''}%
\providecommand \bibnamefont  [1]{#1}%
\providecommand \bibfnamefont [1]{#1}%
\providecommand \citenamefont [1]{#1}%
\providecommand \href@noop [0]{\@secondoftwo}%
\providecommand \href [0]{\begingroup \@sanitize@url \@href}%
\providecommand \@href[1]{\@@startlink{#1}\@@href}%
\providecommand \@@href[1]{\endgroup#1\@@endlink}%
\providecommand \@sanitize@url [0]{\catcode `\\12\catcode `\$12\catcode
  `\&12\catcode `\#12\catcode `\^12\catcode `\_12\catcode `\%12\relax}%
\providecommand \@@startlink[1]{}%
\providecommand \@@endlink[0]{}%
\providecommand \url  [0]{\begingroup\@sanitize@url \@url }%
\providecommand \@url [1]{\endgroup\@href {#1}{\urlprefix }}%
\providecommand \urlprefix  [0]{URL }%
\providecommand \Eprint [0]{\href }%
\providecommand \doibase [0]{http://dx.doi.org/}%
\providecommand \selectlanguage [0]{\@gobble}%
\providecommand \bibinfo  [0]{\@secondoftwo}%
\providecommand \bibfield  [0]{\@secondoftwo}%
\providecommand \translation [1]{[#1]}%
\providecommand \BibitemOpen [0]{}%
\providecommand \bibitemStop [0]{}%
\providecommand \bibitemNoStop [0]{.\EOS\space}%
\providecommand \EOS [0]{\spacefactor3000\relax}%
\providecommand \BibitemShut  [1]{\csname bibitem#1\endcsname}%
\let\auto@bib@innerbib\@empty
\bibitem [{\citenamefont {Fischer}\ \emph {et~al.}(2010)\citenamefont
  {Fischer}, \citenamefont {Whitehouse}, \citenamefont {Mezzacappa},
  \citenamefont {Thielemann},\ and\ \citenamefont {Liebendorfer}}]{fish10}%
  \BibitemOpen
  \bibfield  {author} {\bibinfo {author} {\bibfnamefont {T.}~\bibnamefont
  {Fischer}}, \bibinfo {author} {\bibfnamefont {S.~C.}\ \bibnamefont
  {Whitehouse}}, \bibinfo {author} {\bibfnamefont {A.}~\bibnamefont
  {Mezzacappa}}, \bibinfo {author} {\bibfnamefont {F.~K.}\ \bibnamefont
  {Thielemann}}, \ and\ \bibinfo {author} {\bibfnamefont {M.}~\bibnamefont
  {Liebendorfer}},\ }\href {\doibase 10.1051/0004-6361/200913106} {\bibfield
  {journal} {\bibinfo  {journal} {Astronomy \& Astrophysics}\ }\textbf
  {\bibinfo {volume} {517}},\ \bibinfo {pages} {A80} (\bibinfo {year}
  {2010})}\BibitemShut {NoStop}%
\bibitem [{\citenamefont {Burrows}(2013)}]{burr13}%
  \BibitemOpen
  \bibfield  {author} {\bibinfo {author} {\bibfnamefont {A.}~\bibnamefont
  {Burrows}},\ }\href {\doibase 10.1103/RevModPhys.85.245} {\bibfield
  {journal} {\bibinfo  {journal} {Reviews of Modern Physics}\ }\textbf
  {\bibinfo {volume} {85}},\ \bibinfo {pages} {245} (\bibinfo {year}
  {2013})}\BibitemShut {NoStop}%
\bibitem [{\citenamefont {{Yamada}}\ \emph {et~al.}(2024)\citenamefont
  {{Yamada}}, \citenamefont {{Nagakura}}, \citenamefont {{Akaho}},
  \citenamefont {{Harada}}, \citenamefont {{Furusawa}}, \citenamefont
  {{Iwakami}}, \citenamefont {{Okawa}}, \citenamefont {{Matsufuru}},\ and\
  \citenamefont {{Sumiyoshi}}}]{yama24}%
  \BibitemOpen
  \bibfield  {author} {\bibinfo {author} {\bibfnamefont {S.}~\bibnamefont
  {{Yamada}}}, \bibinfo {author} {\bibfnamefont {H.}~\bibnamefont
  {{Nagakura}}}, \bibinfo {author} {\bibfnamefont {R.}~\bibnamefont {{Akaho}}},
  \bibinfo {author} {\bibfnamefont {A.}~\bibnamefont {{Harada}}}, \bibinfo
  {author} {\bibfnamefont {S.}~\bibnamefont {{Furusawa}}}, \bibinfo {author}
  {\bibfnamefont {W.}~\bibnamefont {{Iwakami}}}, \bibinfo {author}
  {\bibfnamefont {H.}~\bibnamefont {{Okawa}}}, \bibinfo {author} {\bibfnamefont
  {H.}~\bibnamefont {{Matsufuru}}}, \ and\ \bibinfo {author} {\bibfnamefont
  {K.}~\bibnamefont {{Sumiyoshi}}},\ }\href {\doibase 10.2183/pjab.100.015}
  {\bibfield  {journal} {\bibinfo  {journal} {Proceedings of the Japan Academy,
  Series B}\ }\textbf {\bibinfo {volume} {100}},\ \bibinfo {pages} {190}
  (\bibinfo {year} {2024})}\BibitemShut {NoStop}%
\bibitem [{\citenamefont {Sumiyoshi}\ \emph {et~al.}(2019)\citenamefont
  {Sumiyoshi}, \citenamefont {ichiro Nakazato}, \citenamefont {Suzuki},
  \citenamefont {Hu},\ and\ \citenamefont {Shen}}]{sumi19}%
  \BibitemOpen
  \bibfield  {author} {\bibinfo {author} {\bibfnamefont {K.}~\bibnamefont
  {Sumiyoshi}}, \bibinfo {author} {\bibfnamefont {K.}~\bibnamefont {ichiro
  Nakazato}}, \bibinfo {author} {\bibfnamefont {H.}~\bibnamefont {Suzuki}},
  \bibinfo {author} {\bibfnamefont {J.}~\bibnamefont {Hu}}, \ and\ \bibinfo
  {author} {\bibfnamefont {H.}~\bibnamefont {Shen}},\ }\href {\doibase
  10.3847/1538-4357/ab5443} {\bibfield  {journal} {\bibinfo  {journal} {The
  Astrophysical Journal}\ }\textbf {\bibinfo {volume} {887}},\ \bibinfo {pages}
  {110} (\bibinfo {year} {2019})}\BibitemShut {NoStop}%
\bibitem [{\citenamefont {Nakazato}\ and\ \citenamefont
  {Suzuki}(2019)}]{naka19}%
  \BibitemOpen
  \bibfield  {author} {\bibinfo {author} {\bibfnamefont {K.}~\bibnamefont
  {Nakazato}}\ and\ \bibinfo {author} {\bibfnamefont {H.}~\bibnamefont
  {Suzuki}},\ }\href {\doibase 10.3847/1538-4357/ab1d4b} {\bibfield  {journal}
  {\bibinfo  {journal} {The Astrophysical Journal}\ }\textbf {\bibinfo {volume}
  {878}},\ \bibinfo {pages} {25} (\bibinfo {year} {2019})}\BibitemShut
  {NoStop}%
\bibitem [{\citenamefont {Nakazato}\ \emph {et~al.}(2018)\citenamefont
  {Nakazato}, \citenamefont {Suzuki},\ and\ \citenamefont {Togashi}}]{naka18}%
  \BibitemOpen
  \bibfield  {author} {\bibinfo {author} {\bibfnamefont {K.}~\bibnamefont
  {Nakazato}}, \bibinfo {author} {\bibfnamefont {H.}~\bibnamefont {Suzuki}}, \
  and\ \bibinfo {author} {\bibfnamefont {H.}~\bibnamefont {Togashi}},\ }\href
  {\doibase 10.1103/PhysRevC.97.035804} {\bibfield  {journal} {\bibinfo
  {journal} {Physical Review C}\ }\textbf {\bibinfo {volume} {97}},\ \bibinfo
  {pages} {035804} (\bibinfo {year} {2018})}\BibitemShut {NoStop}%
\bibitem [{\citenamefont {Potekhin}\ and\ \citenamefont
  {Chabrier}(2021)}]{pote21}%
  \BibitemOpen
  \bibfield  {author} {\bibinfo {author} {\bibfnamefont {A.~Y.}\ \bibnamefont
  {Potekhin}}\ and\ \bibinfo {author} {\bibfnamefont {G.}~\bibnamefont
  {Chabrier}},\ }\href {\doibase 10.1051/0004-6361/202039006} {\bibfield
  {journal} {\bibinfo  {journal} {Astronomy \& Astrophysics}\ }\textbf
  {\bibinfo {volume} {645}},\ \bibinfo {pages} {A102} (\bibinfo {year}
  {2021})}\BibitemShut {NoStop}%
\bibitem [{\citenamefont {Chamel}\ and\ \citenamefont
  {Haensel}(2008)}]{cham08}%
  \BibitemOpen
  \bibfield  {author} {\bibinfo {author} {\bibfnamefont {N.}~\bibnamefont
  {Chamel}}\ and\ \bibinfo {author} {\bibfnamefont {P.}~\bibnamefont
  {Haensel}},\ }\href {\doibase 10.12942/lrr-2008-10} {\bibfield  {journal}
  {\bibinfo  {journal} {Living Reviews in Relativity}\ }\textbf {\bibinfo
  {volume} {11}},\ \bibinfo {pages} {10} (\bibinfo {year} {2008})}\BibitemShut
  {NoStop}%
\bibitem [{\citenamefont {Lattimer}\ and\ \citenamefont
  {Prakash}(2016)}]{latt16}%
  \BibitemOpen
  \bibfield  {author} {\bibinfo {author} {\bibfnamefont {J.~M.}\ \bibnamefont
  {Lattimer}}\ and\ \bibinfo {author} {\bibfnamefont {M.}~\bibnamefont
  {Prakash}},\ }\href {\doibase https://doi.org/10.1016/j.physrep.2015.12.005}
  {\bibfield  {journal} {\bibinfo  {journal} {Physics Reports}\ }\textbf
  {\bibinfo {volume} {621}},\ \bibinfo {pages} {127} (\bibinfo {year}
  {2016})},\ \bibinfo {note} {memorial Volume in Honor of Gerald E.
  Brown}\BibitemShut {NoStop}%
\bibitem [{\citenamefont {Oertel}\ \emph {et~al.}(2017)\citenamefont {Oertel},
  \citenamefont {Hempel}, \citenamefont {Kl\"ahn},\ and\ \citenamefont
  {Typel}}]{oert17}%
  \BibitemOpen
  \bibfield  {author} {\bibinfo {author} {\bibfnamefont {M.}~\bibnamefont
  {Oertel}}, \bibinfo {author} {\bibfnamefont {M.}~\bibnamefont {Hempel}},
  \bibinfo {author} {\bibfnamefont {T.}~\bibnamefont {Kl\"ahn}}, \ and\
  \bibinfo {author} {\bibfnamefont {S.}~\bibnamefont {Typel}},\ }\href
  {\doibase 10.1103/RevModPhys.89.015007} {\bibfield  {journal} {\bibinfo
  {journal} {Reviews of Modern Physics}\ }\textbf {\bibinfo {volume} {89}},\
  \bibinfo {pages} {015007} (\bibinfo {year} {2017})}\BibitemShut {NoStop}%
\bibitem [{\citenamefont {Koehn}\ \emph {et~al.}(2025)\citenamefont {Koehn},
  \citenamefont {Rose}, \citenamefont {Pang}, \citenamefont {Somasundaram},
  \citenamefont {Reed}, \citenamefont {Tews}, \citenamefont {Abac},
  \citenamefont {Komoltsev}, \citenamefont {Kunert}, \citenamefont {Kurkela},
  \citenamefont {Coughlin}, \citenamefont {Healy},\ and\ \citenamefont
  {Dietrich}}]{koeh25}%
  \BibitemOpen
  \bibfield  {author} {\bibinfo {author} {\bibfnamefont {H.}~\bibnamefont
  {Koehn}}, \bibinfo {author} {\bibfnamefont {H.}~\bibnamefont {Rose}},
  \bibinfo {author} {\bibfnamefont {P.~T.~H.}\ \bibnamefont {Pang}}, \bibinfo
  {author} {\bibfnamefont {R.}~\bibnamefont {Somasundaram}}, \bibinfo {author}
  {\bibfnamefont {B.~T.}\ \bibnamefont {Reed}}, \bibinfo {author}
  {\bibfnamefont {I.}~\bibnamefont {Tews}}, \bibinfo {author} {\bibfnamefont
  {A.}~\bibnamefont {Abac}}, \bibinfo {author} {\bibfnamefont {O.}~\bibnamefont
  {Komoltsev}}, \bibinfo {author} {\bibfnamefont {N.}~\bibnamefont {Kunert}},
  \bibinfo {author} {\bibfnamefont {A.}~\bibnamefont {Kurkela}}, \bibinfo
  {author} {\bibfnamefont {M.~W.}\ \bibnamefont {Coughlin}}, \bibinfo {author}
  {\bibfnamefont {B.~F.}\ \bibnamefont {Healy}}, \ and\ \bibinfo {author}
  {\bibfnamefont {T.}~\bibnamefont {Dietrich}},\ }\href {\doibase
  10.1103/PhysRevX.15.021014} {\bibfield  {journal} {\bibinfo  {journal}
  {Physical Review X}\ }\textbf {\bibinfo {volume} {15}},\ \bibinfo {pages}
  {021014} (\bibinfo {year} {2025})}\BibitemShut {NoStop}%
\bibitem [{\citenamefont {Chatziioannou}\ \emph {et~al.}(2025)\citenamefont
  {Chatziioannou}, \citenamefont {Cromartie}, \citenamefont {Gandolfi},
  \citenamefont {Tews}, \citenamefont {Radice}, \citenamefont {Steiner},\ and\
  \citenamefont {Watts}}]{chat24}%
  \BibitemOpen
  \bibfield  {author} {\bibinfo {author} {\bibfnamefont {K.}~\bibnamefont
  {Chatziioannou}}, \bibinfo {author} {\bibfnamefont {H.~T.}\ \bibnamefont
  {Cromartie}}, \bibinfo {author} {\bibfnamefont {S.}~\bibnamefont {Gandolfi}},
  \bibinfo {author} {\bibfnamefont {I.}~\bibnamefont {Tews}}, \bibinfo {author}
  {\bibfnamefont {D.}~\bibnamefont {Radice}}, \bibinfo {author} {\bibfnamefont
  {A.~W.}\ \bibnamefont {Steiner}}, \ and\ \bibinfo {author} {\bibfnamefont
  {A.~L.}\ \bibnamefont {Watts}},\ }\href {\doibase 10.1103/ymsq-cfcw}
  {\bibfield  {journal} {\bibinfo  {journal} {Reviews of Modern Physics}\
  }\textbf {\bibinfo {volume} {97}},\ \bibinfo {pages} {045007} (\bibinfo
  {year} {2025})}\BibitemShut {NoStop}%
\bibitem [{\citenamefont {Ascenzi}\ \emph {et~al.}(2024)\citenamefont
  {Ascenzi}, \citenamefont {Graber},\ and\ \citenamefont {Rea}}]{asce24}%
  \BibitemOpen
  \bibfield  {author} {\bibinfo {author} {\bibfnamefont {S.}~\bibnamefont
  {Ascenzi}}, \bibinfo {author} {\bibfnamefont {V.}~\bibnamefont {Graber}}, \
  and\ \bibinfo {author} {\bibfnamefont {N.}~\bibnamefont {Rea}},\ }\href
  {\doibase https://doi.org/10.1016/j.astropartphys.2024.102935} {\bibfield
  {journal} {\bibinfo  {journal} {Astroparticle Physics}\ }\textbf {\bibinfo
  {volume} {158}},\ \bibinfo {pages} {102935} (\bibinfo {year}
  {2024})}\BibitemShut {NoStop}%
\bibitem [{\citenamefont {Abbott}\ \emph {et~al.}(2016)\citenamefont {Abbott},
  \citenamefont {Abbott}, \citenamefont {Abbott}, \citenamefont {Abernathy},
  \citenamefont {Acernese}, \citenamefont {Ackley}, \citenamefont {Adams},
  \citenamefont {Adams}, \citenamefont {Addesso} \emph {et~al.}}]{abbo16}%
  \BibitemOpen
  \bibfield  {author} {\bibinfo {author} {\bibfnamefont {B.~P.}\ \bibnamefont
  {Abbott}}, \bibinfo {author} {\bibfnamefont {R.}~\bibnamefont {Abbott}},
  \bibinfo {author} {\bibfnamefont {T.~D.}\ \bibnamefont {Abbott}}, \bibinfo
  {author} {\bibfnamefont {M.~R.}\ \bibnamefont {Abernathy}}, \bibinfo {author}
  {\bibfnamefont {F.}~\bibnamefont {Acernese}}, \bibinfo {author}
  {\bibfnamefont {K.}~\bibnamefont {Ackley}}, \bibinfo {author} {\bibfnamefont
  {C.}~\bibnamefont {Adams}}, \bibinfo {author} {\bibfnamefont
  {T.}~\bibnamefont {Adams}}, \bibinfo {author} {\bibfnamefont
  {P.}~\bibnamefont {Addesso}},  \emph {et~al.} (\bibinfo {collaboration} {LIGO
  Scientific Collaboration and Virgo Collaboration}),\ }\href {\doibase
  10.1103/PhysRevLett.116.061102} {\bibfield  {journal} {\bibinfo  {journal}
  {Physical Review Letters}\ }\textbf {\bibinfo {volume} {116}},\ \bibinfo
  {pages} {061102} (\bibinfo {year} {2016})}\BibitemShut {NoStop}%
\bibitem [{\citenamefont {Abbott}\ \emph {et~al.}(2017)\citenamefont {Abbott},
  \citenamefont {Abbott}, \citenamefont {Abbott}, \citenamefont {Acernese},
  \citenamefont {Ackley}, \citenamefont {Adams}, \citenamefont {Adams},
  \citenamefont {Addesso}, \citenamefont {Adhikari}, \citenamefont {Adya} \emph
  {et~al.}}]{abbo17}%
  \BibitemOpen
  \bibfield  {author} {\bibinfo {author} {\bibfnamefont {B.~P.}\ \bibnamefont
  {Abbott}}, \bibinfo {author} {\bibfnamefont {R.}~\bibnamefont {Abbott}},
  \bibinfo {author} {\bibfnamefont {T.}~\bibnamefont {Abbott}}, \bibinfo
  {author} {\bibfnamefont {F.}~\bibnamefont {Acernese}}, \bibinfo {author}
  {\bibfnamefont {K.}~\bibnamefont {Ackley}}, \bibinfo {author} {\bibfnamefont
  {C.}~\bibnamefont {Adams}}, \bibinfo {author} {\bibfnamefont
  {T.}~\bibnamefont {Adams}}, \bibinfo {author} {\bibfnamefont
  {P.}~\bibnamefont {Addesso}}, \bibinfo {author} {\bibfnamefont
  {R.}~\bibnamefont {Adhikari}}, \bibinfo {author} {\bibfnamefont {V.~B.}\
  \bibnamefont {Adya}},  \emph {et~al.},\ }\href {\doibase
  10.1103/PhysRevLett.119.161101} {\bibfield  {journal} {\bibinfo  {journal}
  {Physical Review Letters}\ }\textbf {\bibinfo {volume} {119}},\ \bibinfo
  {pages} {161101} (\bibinfo {year} {2017})}\BibitemShut {NoStop}%
\bibitem [{\citenamefont {Abbott}\ \emph {et~al.}(2018)\citenamefont {Abbott},
  \citenamefont {Abbott}, \citenamefont {Abbott}, \citenamefont {Acernese},
  \citenamefont {Ackley}, \citenamefont {Adams}, \citenamefont {Adams},
  \citenamefont {Addesso}, \citenamefont {Adhikari}, \citenamefont {Adya} \emph
  {et~al.}}]{abbo18}%
  \BibitemOpen
  \bibfield  {author} {\bibinfo {author} {\bibfnamefont {B.~P.}\ \bibnamefont
  {Abbott}}, \bibinfo {author} {\bibfnamefont {R.}~\bibnamefont {Abbott}},
  \bibinfo {author} {\bibfnamefont {T.}~\bibnamefont {Abbott}}, \bibinfo
  {author} {\bibfnamefont {F.}~\bibnamefont {Acernese}}, \bibinfo {author}
  {\bibfnamefont {K.}~\bibnamefont {Ackley}}, \bibinfo {author} {\bibfnamefont
  {C.}~\bibnamefont {Adams}}, \bibinfo {author} {\bibfnamefont
  {T.}~\bibnamefont {Adams}}, \bibinfo {author} {\bibfnamefont
  {P.}~\bibnamefont {Addesso}}, \bibinfo {author} {\bibfnamefont {R.~X.}\
  \bibnamefont {Adhikari}}, \bibinfo {author} {\bibfnamefont {V.~B.}\
  \bibnamefont {Adya}},  \emph {et~al.},\ }\href {\doibase
  10.1103/PhysRevLett.121.161101} {\bibfield  {journal} {\bibinfo  {journal}
  {Physical Review Letters}\ }\textbf {\bibinfo {volume} {121}},\ \bibinfo
  {pages} {161101} (\bibinfo {year} {2018})}\BibitemShut {NoStop}%
\bibitem [{\citenamefont {Fattoyev}\ \emph {et~al.}(2018)\citenamefont
  {Fattoyev}, \citenamefont {Piekarewicz},\ and\ \citenamefont
  {Horowitz}}]{fatt18}%
  \BibitemOpen
  \bibfield  {author} {\bibinfo {author} {\bibfnamefont {F.~J.}\ \bibnamefont
  {Fattoyev}}, \bibinfo {author} {\bibfnamefont {J.}~\bibnamefont
  {Piekarewicz}}, \ and\ \bibinfo {author} {\bibfnamefont {C.~J.}\ \bibnamefont
  {Horowitz}},\ }\href {\doibase 10.1103/PhysRevLett.120.172702} {\bibfield
  {journal} {\bibinfo  {journal} {Physical Review Letters}\ }\textbf {\bibinfo
  {volume} {120}},\ \bibinfo {pages} {172702} (\bibinfo {year}
  {2018})}\BibitemShut {NoStop}%
\bibitem [{\citenamefont {Most}\ \emph {et~al.}(2018)\citenamefont {Most},
  \citenamefont {Weih}, \citenamefont {Rezzolla},\ and\ \citenamefont
  {Schaffner-Bielich}}]{most18}%
  \BibitemOpen
  \bibfield  {author} {\bibinfo {author} {\bibfnamefont {E.~R.}\ \bibnamefont
  {Most}}, \bibinfo {author} {\bibfnamefont {L.~R.}\ \bibnamefont {Weih}},
  \bibinfo {author} {\bibfnamefont {L.}~\bibnamefont {Rezzolla}}, \ and\
  \bibinfo {author} {\bibfnamefont {J.}~\bibnamefont {Schaffner-Bielich}},\
  }\href {\doibase 10.1103/PhysRevLett.120.261103} {\bibfield  {journal}
  {\bibinfo  {journal} {Physical Review Letters}\ }\textbf {\bibinfo {volume}
  {120}},\ \bibinfo {pages} {261103} (\bibinfo {year} {2018})}\BibitemShut
  {NoStop}%
\bibitem [{\citenamefont {Landry}\ \emph {et~al.}(2020)\citenamefont {Landry},
  \citenamefont {Essick},\ and\ \citenamefont {Chatziioannou}}]{land20}%
  \BibitemOpen
  \bibfield  {author} {\bibinfo {author} {\bibfnamefont {P.}~\bibnamefont
  {Landry}}, \bibinfo {author} {\bibfnamefont {R.}~\bibnamefont {Essick}}, \
  and\ \bibinfo {author} {\bibfnamefont {K.}~\bibnamefont {Chatziioannou}},\
  }\href {\doibase 10.1103/PhysRevD.101.123007} {\bibfield  {journal} {\bibinfo
   {journal} {Physical Review D}\ }\textbf {\bibinfo {volume} {101}},\ \bibinfo
  {pages} {123007} (\bibinfo {year} {2020})}\BibitemShut {NoStop}%
\bibitem [{\citenamefont {Chatziioannou}(2020)}]{chat20}%
  \BibitemOpen
  \bibfield  {author} {\bibinfo {author} {\bibfnamefont {K.}~\bibnamefont
  {Chatziioannou}},\ }\href {\doibase 10.1007/s10714-020-02754-3} {\bibfield
  {journal} {\bibinfo  {journal} {General Relativity and Gravitation}\ }\textbf
  {\bibinfo {volume} {52}},\ \bibinfo {pages} {109} (\bibinfo {year}
  {2020})}\BibitemShut {NoStop}%
\bibitem [{\citenamefont {Demorest}\ \emph {et~al.}(2010)\citenamefont
  {Demorest}, \citenamefont {Pennucci}, \citenamefont {Ransom}, \citenamefont
  {Roberts},\ and\ \citenamefont {Hessels}}]{demo10}%
  \BibitemOpen
  \bibfield  {author} {\bibinfo {author} {\bibfnamefont {P.~B.}\ \bibnamefont
  {Demorest}}, \bibinfo {author} {\bibfnamefont {T.}~\bibnamefont {Pennucci}},
  \bibinfo {author} {\bibfnamefont {S.}~\bibnamefont {Ransom}}, \bibinfo
  {author} {\bibfnamefont {M.}~\bibnamefont {Roberts}}, \ and\ \bibinfo
  {author} {\bibfnamefont {J.}~\bibnamefont {Hessels}},\ }\href {\doibase
  https://doi.org/10.1038/nature09466} {\bibfield  {journal} {\bibinfo
  {journal} {Nature}\ }\textbf {\bibinfo {volume} {467}},\ \bibinfo {pages}
  {1081} (\bibinfo {year} {2010})}\BibitemShut {NoStop}%
\bibitem [{\citenamefont {Arzoumanian}\ \emph {et~al.}(2018)\citenamefont
  {Arzoumanian}, \citenamefont {Brazier}, \citenamefont {Burke-Spolaor},
  \citenamefont {Chamberlin}, \citenamefont {Chatterjee}, \citenamefont
  {Christy}, \citenamefont {Cordes}, \citenamefont {Cornish}, \citenamefont
  {Crawford}, \citenamefont {Cromartie}, \citenamefont {Crowter}, \citenamefont
  {DeCesar}, \citenamefont {Demorest}, \citenamefont {Dolch}, \citenamefont
  {Ellis}, \citenamefont {Ferdman}, \citenamefont {Ferrara}, \citenamefont
  {Fonseca}, \citenamefont {Garver-Daniels}, \citenamefont {Gentile},
  \citenamefont {Halmrast}, \citenamefont {Huerta}, \citenamefont {Jenet},
  \citenamefont {Jessup}, \citenamefont {Jones}, \citenamefont {Jones},
  \citenamefont {Kaplan}, \citenamefont {Lam}, \citenamefont {Lazio},
  \citenamefont {Levin}, \citenamefont {Lommen}, \citenamefont {Lorimer},
  \citenamefont {Luo}, \citenamefont {Lynch}, \citenamefont {Madison},
  \citenamefont {Matthews}, \citenamefont {McLaughlin}, \citenamefont
  {McWilliams}, \citenamefont {Mingarelli}, \citenamefont {Ng}, \citenamefont
  {Nice}, \citenamefont {Pennucci}, \citenamefont {Ransom}, \citenamefont
  {Ray}, \citenamefont {Siemens}, \citenamefont {Simon}, \citenamefont
  {Spiewak}, \citenamefont {Stairs}, \citenamefont {Stinebring}, \citenamefont
  {Stovall}, \citenamefont {Swiggum}, \citenamefont {Taylor}, \citenamefont
  {Vallisneri}, \citenamefont {van Haasteren}, \citenamefont {Vigeland},
  \citenamefont {Zhu},\ and\ \citenamefont {Collaboration}}]{arzo18}%
  \BibitemOpen
  \bibfield  {author} {\bibinfo {author} {\bibfnamefont {Z.}~\bibnamefont
  {Arzoumanian}}, \bibinfo {author} {\bibfnamefont {A.}~\bibnamefont
  {Brazier}}, \bibinfo {author} {\bibfnamefont {S.}~\bibnamefont
  {Burke-Spolaor}}, \bibinfo {author} {\bibfnamefont {S.}~\bibnamefont
  {Chamberlin}}, \bibinfo {author} {\bibfnamefont {S.}~\bibnamefont
  {Chatterjee}}, \bibinfo {author} {\bibfnamefont {B.}~\bibnamefont {Christy}},
  \bibinfo {author} {\bibfnamefont {J.~M.}\ \bibnamefont {Cordes}}, \bibinfo
  {author} {\bibfnamefont {N.~J.}\ \bibnamefont {Cornish}}, \bibinfo {author}
  {\bibfnamefont {F.}~\bibnamefont {Crawford}}, \bibinfo {author}
  {\bibfnamefont {H.~T.}\ \bibnamefont {Cromartie}}, \bibinfo {author}
  {\bibfnamefont {K.}~\bibnamefont {Crowter}}, \bibinfo {author} {\bibfnamefont
  {M.~E.}\ \bibnamefont {DeCesar}}, \bibinfo {author} {\bibfnamefont {P.~B.}\
  \bibnamefont {Demorest}}, \bibinfo {author} {\bibfnamefont {T.}~\bibnamefont
  {Dolch}}, \bibinfo {author} {\bibfnamefont {J.~A.}\ \bibnamefont {Ellis}},
  \bibinfo {author} {\bibfnamefont {R.~D.}\ \bibnamefont {Ferdman}}, \bibinfo
  {author} {\bibfnamefont {E.~C.}\ \bibnamefont {Ferrara}}, \bibinfo {author}
  {\bibfnamefont {E.}~\bibnamefont {Fonseca}}, \bibinfo {author} {\bibfnamefont
  {N.}~\bibnamefont {Garver-Daniels}}, \bibinfo {author} {\bibfnamefont
  {P.~A.}\ \bibnamefont {Gentile}}, \bibinfo {author} {\bibfnamefont
  {D.}~\bibnamefont {Halmrast}}, \bibinfo {author} {\bibfnamefont {E.~A.}\
  \bibnamefont {Huerta}}, \bibinfo {author} {\bibfnamefont {F.~A.}\
  \bibnamefont {Jenet}}, \bibinfo {author} {\bibfnamefont {C.}~\bibnamefont
  {Jessup}}, \bibinfo {author} {\bibfnamefont {G.}~\bibnamefont {Jones}},
  \bibinfo {author} {\bibfnamefont {M.~L.}\ \bibnamefont {Jones}}, \bibinfo
  {author} {\bibfnamefont {D.~L.}\ \bibnamefont {Kaplan}}, \bibinfo {author}
  {\bibfnamefont {M.~T.}\ \bibnamefont {Lam}}, \bibinfo {author} {\bibfnamefont
  {T.~J.~W.}\ \bibnamefont {Lazio}}, \bibinfo {author} {\bibfnamefont
  {L.}~\bibnamefont {Levin}}, \bibinfo {author} {\bibfnamefont
  {A.}~\bibnamefont {Lommen}}, \bibinfo {author} {\bibfnamefont {D.~R.}\
  \bibnamefont {Lorimer}}, \bibinfo {author} {\bibfnamefont {J.}~\bibnamefont
  {Luo}}, \bibinfo {author} {\bibfnamefont {R.~S.}\ \bibnamefont {Lynch}},
  \bibinfo {author} {\bibfnamefont {D.}~\bibnamefont {Madison}}, \bibinfo
  {author} {\bibfnamefont {A.~M.}\ \bibnamefont {Matthews}}, \bibinfo {author}
  {\bibfnamefont {M.~A.}\ \bibnamefont {McLaughlin}}, \bibinfo {author}
  {\bibfnamefont {S.~T.}\ \bibnamefont {McWilliams}}, \bibinfo {author}
  {\bibfnamefont {C.}~\bibnamefont {Mingarelli}}, \bibinfo {author}
  {\bibfnamefont {C.}~\bibnamefont {Ng}}, \bibinfo {author} {\bibfnamefont
  {D.~J.}\ \bibnamefont {Nice}}, \bibinfo {author} {\bibfnamefont {T.~T.}\
  \bibnamefont {Pennucci}}, \bibinfo {author} {\bibfnamefont {S.~M.}\
  \bibnamefont {Ransom}}, \bibinfo {author} {\bibfnamefont {P.~S.}\
  \bibnamefont {Ray}}, \bibinfo {author} {\bibfnamefont {X.}~\bibnamefont
  {Siemens}}, \bibinfo {author} {\bibfnamefont {J.}~\bibnamefont {Simon}},
  \bibinfo {author} {\bibfnamefont {R.}~\bibnamefont {Spiewak}}, \bibinfo
  {author} {\bibfnamefont {I.~H.}\ \bibnamefont {Stairs}}, \bibinfo {author}
  {\bibfnamefont {D.~R.}\ \bibnamefont {Stinebring}}, \bibinfo {author}
  {\bibfnamefont {K.}~\bibnamefont {Stovall}}, \bibinfo {author} {\bibfnamefont
  {J.~K.}\ \bibnamefont {Swiggum}}, \bibinfo {author} {\bibfnamefont {S.~R.}\
  \bibnamefont {Taylor}}, \bibinfo {author} {\bibfnamefont {M.}~\bibnamefont
  {Vallisneri}}, \bibinfo {author} {\bibfnamefont {R.}~\bibnamefont {van
  Haasteren}}, \bibinfo {author} {\bibfnamefont {S.~J.}\ \bibnamefont
  {Vigeland}}, \bibinfo {author} {\bibfnamefont {W.}~\bibnamefont {Zhu}}, \
  and\ \bibinfo {author} {\bibfnamefont {T.~N.}\ \bibnamefont
  {Collaboration}},\ }\href {\doibase 10.3847/1538-4365/aab5b0} {\bibfield
  {journal} {\bibinfo  {journal} {The Astrophysical Journal Supplement Series}\
  }\textbf {\bibinfo {volume} {235}},\ \bibinfo {pages} {37} (\bibinfo {year}
  {2018})}\BibitemShut {NoStop}%
\bibitem [{\citenamefont {Antoniadis}\ \emph {et~al.}(2013)\citenamefont
  {Antoniadis}, \citenamefont {Freire}, \citenamefont {Wex}, \citenamefont
  {Tauris}, \citenamefont {Lynch}, \citenamefont {van Kerkwijk}, \citenamefont
  {Kramer}, \citenamefont {Bassa}, \citenamefont {Dhillon}, \citenamefont
  {Driebe}, \citenamefont {Hessels}, \citenamefont {Kaspi}, \citenamefont
  {Kondratiev}, \citenamefont {Langer}, \citenamefont {Marsh}, \citenamefont
  {McLaughlin}, \citenamefont {Pennucci}, \citenamefont {Ransom}, \citenamefont
  {Stairs}, \citenamefont {van Leeuwen}, \citenamefont {Verbiest},\ and\
  \citenamefont {Whelan}}]{anto13}%
  \BibitemOpen
  \bibfield  {author} {\bibinfo {author} {\bibfnamefont {J.}~\bibnamefont
  {Antoniadis}}, \bibinfo {author} {\bibfnamefont {P.~C.~C.}\ \bibnamefont
  {Freire}}, \bibinfo {author} {\bibfnamefont {N.}~\bibnamefont {Wex}},
  \bibinfo {author} {\bibfnamefont {T.~M.}\ \bibnamefont {Tauris}}, \bibinfo
  {author} {\bibfnamefont {R.~S.}\ \bibnamefont {Lynch}}, \bibinfo {author}
  {\bibfnamefont {M.~H.}\ \bibnamefont {van Kerkwijk}}, \bibinfo {author}
  {\bibfnamefont {M.}~\bibnamefont {Kramer}}, \bibinfo {author} {\bibfnamefont
  {C.}~\bibnamefont {Bassa}}, \bibinfo {author} {\bibfnamefont {V.~S.}\
  \bibnamefont {Dhillon}}, \bibinfo {author} {\bibfnamefont {T.}~\bibnamefont
  {Driebe}}, \bibinfo {author} {\bibfnamefont {J.~W.~T.}\ \bibnamefont
  {Hessels}}, \bibinfo {author} {\bibfnamefont {V.~M.}\ \bibnamefont {Kaspi}},
  \bibinfo {author} {\bibfnamefont {V.~I.}\ \bibnamefont {Kondratiev}},
  \bibinfo {author} {\bibfnamefont {N.}~\bibnamefont {Langer}}, \bibinfo
  {author} {\bibfnamefont {T.~R.}\ \bibnamefont {Marsh}}, \bibinfo {author}
  {\bibfnamefont {M.~A.}\ \bibnamefont {McLaughlin}}, \bibinfo {author}
  {\bibfnamefont {T.~T.}\ \bibnamefont {Pennucci}}, \bibinfo {author}
  {\bibfnamefont {S.~M.}\ \bibnamefont {Ransom}}, \bibinfo {author}
  {\bibfnamefont {I.~H.}\ \bibnamefont {Stairs}}, \bibinfo {author}
  {\bibfnamefont {J.}~\bibnamefont {van Leeuwen}}, \bibinfo {author}
  {\bibfnamefont {J.~P.~W.}\ \bibnamefont {Verbiest}}, \ and\ \bibinfo {author}
  {\bibfnamefont {D.~G.}\ \bibnamefont {Whelan}},\ }\href {\doibase
  10.1126/science.1233232} {\bibfield  {journal} {\bibinfo  {journal}
  {Science}\ }\textbf {\bibinfo {volume} {340}},\ \bibinfo {pages} {1233232}
  (\bibinfo {year} {2013})}\BibitemShut {NoStop}%
\bibitem [{\citenamefont {Fonseca}\ \emph {et~al.}(2021)\citenamefont
  {Fonseca}, \citenamefont {Cromartie}, \citenamefont {Pennucci}, \citenamefont
  {Ray}, \citenamefont {Kirichenko}, \citenamefont {Ransom}, \citenamefont
  {Demorest}, \citenamefont {Stairs}, \citenamefont {Arzoumanian},
  \citenamefont {Guillemot}, \citenamefont {Parthasarathy}, \citenamefont
  {Kerr}, \citenamefont {Cognard}, \citenamefont {Baker}, \citenamefont
  {Blumer}, \citenamefont {Brook}, \citenamefont {DeCesar}, \citenamefont
  {Dolch}, \citenamefont {Dong}, \citenamefont {Ferrara}, \citenamefont
  {Fiore}, \citenamefont {Garver-Daniels}, \citenamefont {Good}, \citenamefont
  {Jennings}, \citenamefont {Jones}, \citenamefont {Kaspi}, \citenamefont
  {Lam}, \citenamefont {Lorimer}, \citenamefont {Luo}, \citenamefont {McEwen},
  \citenamefont {McKee}, \citenamefont {McLaughlin}, \citenamefont {McMann},
  \citenamefont {Meyers}, \citenamefont {Naidu}, \citenamefont {Ng},
  \citenamefont {Nice}, \citenamefont {Pol}, \citenamefont {Radovan},
  \citenamefont {Shapiro-Albert}, \citenamefont {Tan}, \citenamefont
  {Tendulkar}, \citenamefont {Swiggum}, \citenamefont {Wahl},\ and\
  \citenamefont {Zhu}}]{fons21}%
  \BibitemOpen
  \bibfield  {author} {\bibinfo {author} {\bibfnamefont {E.}~\bibnamefont
  {Fonseca}}, \bibinfo {author} {\bibfnamefont {H.~T.}\ \bibnamefont
  {Cromartie}}, \bibinfo {author} {\bibfnamefont {T.~T.}\ \bibnamefont
  {Pennucci}}, \bibinfo {author} {\bibfnamefont {P.~S.}\ \bibnamefont {Ray}},
  \bibinfo {author} {\bibfnamefont {A.~Y.}\ \bibnamefont {Kirichenko}},
  \bibinfo {author} {\bibfnamefont {S.~M.}\ \bibnamefont {Ransom}}, \bibinfo
  {author} {\bibfnamefont {P.~B.}\ \bibnamefont {Demorest}}, \bibinfo {author}
  {\bibfnamefont {I.~H.}\ \bibnamefont {Stairs}}, \bibinfo {author}
  {\bibfnamefont {Z.}~\bibnamefont {Arzoumanian}}, \bibinfo {author}
  {\bibfnamefont {L.}~\bibnamefont {Guillemot}}, \bibinfo {author}
  {\bibfnamefont {A.}~\bibnamefont {Parthasarathy}}, \bibinfo {author}
  {\bibfnamefont {M.}~\bibnamefont {Kerr}}, \bibinfo {author} {\bibfnamefont
  {I.}~\bibnamefont {Cognard}}, \bibinfo {author} {\bibfnamefont {P.~T.}\
  \bibnamefont {Baker}}, \bibinfo {author} {\bibfnamefont {H.}~\bibnamefont
  {Blumer}}, \bibinfo {author} {\bibfnamefont {P.~R.}\ \bibnamefont {Brook}},
  \bibinfo {author} {\bibfnamefont {M.}~\bibnamefont {DeCesar}}, \bibinfo
  {author} {\bibfnamefont {T.}~\bibnamefont {Dolch}}, \bibinfo {author}
  {\bibfnamefont {F.~A.}\ \bibnamefont {Dong}}, \bibinfo {author}
  {\bibfnamefont {E.~C.}\ \bibnamefont {Ferrara}}, \bibinfo {author}
  {\bibfnamefont {W.}~\bibnamefont {Fiore}}, \bibinfo {author} {\bibfnamefont
  {N.}~\bibnamefont {Garver-Daniels}}, \bibinfo {author} {\bibfnamefont
  {D.~C.}\ \bibnamefont {Good}}, \bibinfo {author} {\bibfnamefont
  {R.}~\bibnamefont {Jennings}}, \bibinfo {author} {\bibfnamefont {M.~L.}\
  \bibnamefont {Jones}}, \bibinfo {author} {\bibfnamefont {V.~M.}\ \bibnamefont
  {Kaspi}}, \bibinfo {author} {\bibfnamefont {M.~T.}\ \bibnamefont {Lam}},
  \bibinfo {author} {\bibfnamefont {D.~R.}\ \bibnamefont {Lorimer}}, \bibinfo
  {author} {\bibfnamefont {J.}~\bibnamefont {Luo}}, \bibinfo {author}
  {\bibfnamefont {A.}~\bibnamefont {McEwen}}, \bibinfo {author} {\bibfnamefont
  {J.~W.}\ \bibnamefont {McKee}}, \bibinfo {author} {\bibfnamefont {M.~A.}\
  \bibnamefont {McLaughlin}}, \bibinfo {author} {\bibfnamefont
  {N.}~\bibnamefont {McMann}}, \bibinfo {author} {\bibfnamefont {B.~W.}\
  \bibnamefont {Meyers}}, \bibinfo {author} {\bibfnamefont {A.}~\bibnamefont
  {Naidu}}, \bibinfo {author} {\bibfnamefont {C.}~\bibnamefont {Ng}}, \bibinfo
  {author} {\bibfnamefont {D.~J.}\ \bibnamefont {Nice}}, \bibinfo {author}
  {\bibfnamefont {N.}~\bibnamefont {Pol}}, \bibinfo {author} {\bibfnamefont
  {H.~A.}\ \bibnamefont {Radovan}}, \bibinfo {author} {\bibfnamefont
  {B.}~\bibnamefont {Shapiro-Albert}}, \bibinfo {author} {\bibfnamefont
  {C.~M.}\ \bibnamefont {Tan}}, \bibinfo {author} {\bibfnamefont {S.~P.}\
  \bibnamefont {Tendulkar}}, \bibinfo {author} {\bibfnamefont {J.~K.}\
  \bibnamefont {Swiggum}}, \bibinfo {author} {\bibfnamefont {H.~M.}\
  \bibnamefont {Wahl}}, \ and\ \bibinfo {author} {\bibfnamefont {W.~W.}\
  \bibnamefont {Zhu}},\ }\href {\doibase 10.3847/2041-8213/ac03b8} {\bibfield
  {journal} {\bibinfo  {journal} {The Astrophysical Journal Letters}\ }\textbf
  {\bibinfo {volume} {915}},\ \bibinfo {pages} {L12} (\bibinfo {year}
  {2021})}\BibitemShut {NoStop}%
\bibitem [{\citenamefont {Miller}\ \emph {et~al.}(2019)\citenamefont {Miller},
  \citenamefont {Lamb}, \citenamefont {Dittmann}, \citenamefont {Bogdanov},
  \citenamefont {Arzoumanian}, \citenamefont {Gendreau}, \citenamefont
  {Guillot}, \citenamefont {Harding}, \citenamefont {Ho}, \citenamefont
  {Lattimer} \emph {et~al.}}]{mill19}%
  \BibitemOpen
  \bibfield  {author} {\bibinfo {author} {\bibfnamefont {M.~C.}\ \bibnamefont
  {Miller}}, \bibinfo {author} {\bibfnamefont {F.~K.}\ \bibnamefont {Lamb}},
  \bibinfo {author} {\bibfnamefont {A.~J.}\ \bibnamefont {Dittmann}}, \bibinfo
  {author} {\bibfnamefont {S.}~\bibnamefont {Bogdanov}}, \bibinfo {author}
  {\bibfnamefont {Z.}~\bibnamefont {Arzoumanian}}, \bibinfo {author}
  {\bibfnamefont {K.~C.}\ \bibnamefont {Gendreau}}, \bibinfo {author}
  {\bibfnamefont {S.}~\bibnamefont {Guillot}}, \bibinfo {author} {\bibfnamefont
  {A.}~\bibnamefont {Harding}}, \bibinfo {author} {\bibfnamefont
  {W.}~\bibnamefont {Ho}}, \bibinfo {author} {\bibfnamefont {J.~M.}\
  \bibnamefont {Lattimer}},  \emph {et~al.},\ }\href {\doibase
  10.3847/2041-8213/ab50c5} {\bibfield  {journal} {\bibinfo  {journal} {The
  Astrophysical Journal Letters}\ }\textbf {\bibinfo {volume} {887}},\ \bibinfo
  {pages} {L24} (\bibinfo {year} {2019})}\BibitemShut {NoStop}%
\bibitem [{\citenamefont {Riley}\ \emph {et~al.}(2019)\citenamefont {Riley},
  \citenamefont {Watts}, \citenamefont {Bogdanov}, \citenamefont {Ray},
  \citenamefont {Ludlam}, \citenamefont {Guillot}, \citenamefont {Arzoumanian},
  \citenamefont {Baker}, \citenamefont {Bilous}, \citenamefont {Chakrabarty}
  \emph {et~al.}}]{rile19}%
  \BibitemOpen
  \bibfield  {author} {\bibinfo {author} {\bibfnamefont {T.~E.}\ \bibnamefont
  {Riley}}, \bibinfo {author} {\bibfnamefont {A.~L.}\ \bibnamefont {Watts}},
  \bibinfo {author} {\bibfnamefont {S.}~\bibnamefont {Bogdanov}}, \bibinfo
  {author} {\bibfnamefont {P.~S.}\ \bibnamefont {Ray}}, \bibinfo {author}
  {\bibfnamefont {R.~M.}\ \bibnamefont {Ludlam}}, \bibinfo {author}
  {\bibfnamefont {S.}~\bibnamefont {Guillot}}, \bibinfo {author} {\bibfnamefont
  {Z.}~\bibnamefont {Arzoumanian}}, \bibinfo {author} {\bibfnamefont {C.~L.}\
  \bibnamefont {Baker}}, \bibinfo {author} {\bibfnamefont {A.~V.}\ \bibnamefont
  {Bilous}}, \bibinfo {author} {\bibfnamefont {D.}~\bibnamefont {Chakrabarty}},
   \emph {et~al.},\ }\href {\doibase 10.3847/2041-8213/ab481c} {\bibfield
  {journal} {\bibinfo  {journal} {The Astrophysical Journal Letters}\ }\textbf
  {\bibinfo {volume} {887}},\ \bibinfo {pages} {L21} (\bibinfo {year}
  {2019})}\BibitemShut {NoStop}%
\bibitem [{\citenamefont {Miller}\ \emph {et~al.}(2021)\citenamefont {Miller},
  \citenamefont {Lamb}, \citenamefont {Dittmann}, \citenamefont {Bogdanov},
  \citenamefont {Arzoumanian}, \citenamefont {Gendreau}, \citenamefont
  {Guillot}, \citenamefont {Ho}, \citenamefont {Lattimer}, \citenamefont
  {Loewenstein} \emph {et~al.}}]{mill21}%
  \BibitemOpen
  \bibfield  {author} {\bibinfo {author} {\bibfnamefont {M.~C.}\ \bibnamefont
  {Miller}}, \bibinfo {author} {\bibfnamefont {F.~K.}\ \bibnamefont {Lamb}},
  \bibinfo {author} {\bibfnamefont {A.~J.}\ \bibnamefont {Dittmann}}, \bibinfo
  {author} {\bibfnamefont {S.}~\bibnamefont {Bogdanov}}, \bibinfo {author}
  {\bibfnamefont {Z.}~\bibnamefont {Arzoumanian}}, \bibinfo {author}
  {\bibfnamefont {K.~C.}\ \bibnamefont {Gendreau}}, \bibinfo {author}
  {\bibfnamefont {S.}~\bibnamefont {Guillot}}, \bibinfo {author} {\bibfnamefont
  {W.~C.~G.}\ \bibnamefont {Ho}}, \bibinfo {author} {\bibfnamefont {J.~M.}\
  \bibnamefont {Lattimer}}, \bibinfo {author} {\bibfnamefont {M.}~\bibnamefont
  {Loewenstein}},  \emph {et~al.},\ }\href {\doibase 10.3847/2041-8213/ac089b}
  {\bibfield  {journal} {\bibinfo  {journal} {The Astrophysical Journal
  Letters}\ }\textbf {\bibinfo {volume} {918}},\ \bibinfo {pages} {L28}
  (\bibinfo {year} {2021})}\BibitemShut {NoStop}%
\bibitem [{\citenamefont {Riley}\ \emph {et~al.}(2021)\citenamefont {Riley},
  \citenamefont {Watts}, \citenamefont {Ray}, \citenamefont {Bogdanov},
  \citenamefont {Guillot}, \citenamefont {Morsink}, \citenamefont {Bilous},
  \citenamefont {Arzoumanian}, \citenamefont {Choudhury}, \citenamefont
  {Deneva} \emph {et~al.}}]{rile21}%
  \BibitemOpen
  \bibfield  {author} {\bibinfo {author} {\bibfnamefont {T.~E.}\ \bibnamefont
  {Riley}}, \bibinfo {author} {\bibfnamefont {A.~L.}\ \bibnamefont {Watts}},
  \bibinfo {author} {\bibfnamefont {P.~S.}\ \bibnamefont {Ray}}, \bibinfo
  {author} {\bibfnamefont {S.}~\bibnamefont {Bogdanov}}, \bibinfo {author}
  {\bibfnamefont {S.}~\bibnamefont {Guillot}}, \bibinfo {author} {\bibfnamefont
  {S.~M.}\ \bibnamefont {Morsink}}, \bibinfo {author} {\bibfnamefont {A.~V.}\
  \bibnamefont {Bilous}}, \bibinfo {author} {\bibfnamefont {Z.}~\bibnamefont
  {Arzoumanian}}, \bibinfo {author} {\bibfnamefont {D.}~\bibnamefont
  {Choudhury}}, \bibinfo {author} {\bibfnamefont {J.~S.}\ \bibnamefont
  {Deneva}},  \emph {et~al.},\ }\href {\doibase 10.3847/2041-8213/ac0a81}
  {\bibfield  {journal} {\bibinfo  {journal} {The Astrophysical Journal
  Letters}\ }\textbf {\bibinfo {volume} {918}},\ \bibinfo {pages} {L27}
  (\bibinfo {year} {2021})}\BibitemShut {NoStop}%
\bibitem [{\citenamefont {Dutra}\ \emph {et~al.}(2014)\citenamefont {Dutra},
  \citenamefont {Louren{\c{c}}o}, \citenamefont {Avancini}, \citenamefont
  {Carlson}, \citenamefont {Delfino}, \citenamefont {Menezes}, \citenamefont
  {Provid{\^e}ncia}, \citenamefont {Typel},\ and\ \citenamefont
  {Stone}}]{dutr14}%
  \BibitemOpen
  \bibfield  {author} {\bibinfo {author} {\bibfnamefont {M.}~\bibnamefont
  {Dutra}}, \bibinfo {author} {\bibfnamefont {O.}~\bibnamefont
  {Louren{\c{c}}o}}, \bibinfo {author} {\bibfnamefont {S.}~\bibnamefont
  {Avancini}}, \bibinfo {author} {\bibfnamefont {B.~V.}\ \bibnamefont
  {Carlson}}, \bibinfo {author} {\bibfnamefont {A.}~\bibnamefont {Delfino}},
  \bibinfo {author} {\bibfnamefont {D.}~\bibnamefont {Menezes}}, \bibinfo
  {author} {\bibfnamefont {C.}~\bibnamefont {Provid{\^e}ncia}}, \bibinfo
  {author} {\bibfnamefont {S.}~\bibnamefont {Typel}}, \ and\ \bibinfo {author}
  {\bibfnamefont {J.}~\bibnamefont {Stone}},\ }\href@noop {} {\bibfield
  {journal} {\bibinfo  {journal} {Physical Review C}\ }\textbf {\bibinfo
  {volume} {90}},\ \bibinfo {pages} {055203} (\bibinfo {year}
  {2014})}\BibitemShut {NoStop}%
\bibitem [{\citenamefont {Miyatsu}\ \emph {et~al.}(2025)\citenamefont
  {Miyatsu}, \citenamefont {Cheoun}, \citenamefont {Kim},\ and\ \citenamefont
  {Saito}}]{miya25}%
  \BibitemOpen
  \bibfield  {author} {\bibinfo {author} {\bibfnamefont {T.}~\bibnamefont
  {Miyatsu}}, \bibinfo {author} {\bibfnamefont {M.-K.}\ \bibnamefont {Cheoun}},
  \bibinfo {author} {\bibfnamefont {K.}~\bibnamefont {Kim}}, \ and\ \bibinfo
  {author} {\bibfnamefont {K.}~\bibnamefont {Saito}},\ }\href {\doibase
  10.3389/fphy.2024.1531475} {\bibfield  {journal} {\bibinfo  {journal}
  {Frontiers in Physics}\ }\textbf {\bibinfo {volume} {12}},\ \bibinfo {pages}
  {1531475} (\bibinfo {year} {2025})}\BibitemShut {NoStop}%
\bibitem [{\citenamefont {Editors}(2011)}]{PRA11}%
  \BibitemOpen
  \bibfield  {author} {\bibinfo {author} {\bibfnamefont {T.}~\bibnamefont
  {Editors}},\ }\href {\doibase 10.1103/PhysRevA.83.040001} {\bibfield
  {journal} {\bibinfo  {journal} {Phys. Rev. A}\ }\textbf {\bibinfo {volume}
  {83}},\ \bibinfo {pages} {040001} (\bibinfo {year} {2011})}\BibitemShut
  {NoStop}%
\bibitem [{\citenamefont {Chen}\ and\ \citenamefont
  {Piekarewicz}(2014)}]{chen14}%
  \BibitemOpen
  \bibfield  {author} {\bibinfo {author} {\bibfnamefont {W.-C.}\ \bibnamefont
  {Chen}}\ and\ \bibinfo {author} {\bibfnamefont {J.}~\bibnamefont
  {Piekarewicz}},\ }\href {\doibase 10.1103/PhysRevC.90.044305} {\bibfield
  {journal} {\bibinfo  {journal} {Physical Review C}\ }\textbf {\bibinfo
  {volume} {90}},\ \bibinfo {pages} {044305} (\bibinfo {year}
  {2014})}\BibitemShut {NoStop}%
\bibitem [{\citenamefont {Dobaczewski}\ \emph {et~al.}(2014)\citenamefont
  {Dobaczewski}, \citenamefont {Nazarewicz},\ and\ \citenamefont
  {Reinhard}}]{doba14}%
  \BibitemOpen
  \bibfield  {author} {\bibinfo {author} {\bibfnamefont {J.}~\bibnamefont
  {Dobaczewski}}, \bibinfo {author} {\bibfnamefont {W.}~\bibnamefont
  {Nazarewicz}}, \ and\ \bibinfo {author} {\bibfnamefont {P.-G.}\ \bibnamefont
  {Reinhard}},\ }\href {\doibase 10.1088/0954-3899/41/7/074001} {\bibfield
  {journal} {\bibinfo  {journal} {Journal of Physics G: Nuclear and Particle
  Physics}\ }\textbf {\bibinfo {volume} {41}},\ \bibinfo {pages} {074001}
  (\bibinfo {year} {2014})}\BibitemShut {NoStop}%
\bibitem [{\citenamefont {Yang}\ and\ \citenamefont
  {Piekarewicz}(2020)}]{yang20}%
  \BibitemOpen
  \bibfield  {author} {\bibinfo {author} {\bibfnamefont {J.}~\bibnamefont
  {Yang}}\ and\ \bibinfo {author} {\bibfnamefont {J.}~\bibnamefont
  {Piekarewicz}},\ }\href {\doibase
  https://doi.org/10.1146/annurev-nucl-101918-023608} {\bibfield  {journal}
  {\bibinfo  {journal} {Annual Review of Nuclear and Particle Science}\
  }\textbf {\bibinfo {volume} {70}},\ \bibinfo {pages} {21} (\bibinfo {year}
  {2020})}\BibitemShut {NoStop}%
\bibitem [{\citenamefont {Sugahara}\ and\ \citenamefont {Toki}(1994)}]{suga94}%
  \BibitemOpen
  \bibfield  {author} {\bibinfo {author} {\bibfnamefont {Y.}~\bibnamefont
  {Sugahara}}\ and\ \bibinfo {author} {\bibfnamefont {H.}~\bibnamefont
  {Toki}},\ }\href {\doibase https://doi.org/10.1016/0375-9474(94)90923-7}
  {\bibfield  {journal} {\bibinfo  {journal} {Nuclear Physics A}\ }\textbf
  {\bibinfo {volume} {579}},\ \bibinfo {pages} {557} (\bibinfo {year}
  {1994})}\BibitemShut {NoStop}%
\bibitem [{\citenamefont {Fattoyev}\ \emph {et~al.}(2010)\citenamefont
  {Fattoyev}, \citenamefont {Horowitz}, \citenamefont {Piekarewicz},\ and\
  \citenamefont {Shen}}]{fatt10}%
  \BibitemOpen
  \bibfield  {author} {\bibinfo {author} {\bibfnamefont {F.~J.}\ \bibnamefont
  {Fattoyev}}, \bibinfo {author} {\bibfnamefont {C.~J.}\ \bibnamefont
  {Horowitz}}, \bibinfo {author} {\bibfnamefont {J.}~\bibnamefont
  {Piekarewicz}}, \ and\ \bibinfo {author} {\bibfnamefont {G.}~\bibnamefont
  {Shen}},\ }\href {\doibase 10.1103/PhysRevC.82.055803} {\bibfield  {journal}
  {\bibinfo  {journal} {Physical Review C}\ }\textbf {\bibinfo {volume} {82}},\
  \bibinfo {pages} {055803} (\bibinfo {year} {2010})}\BibitemShut {NoStop}%
\bibitem [{\citenamefont {Bao}\ \emph {et~al.}(2014)\citenamefont {Bao},
  \citenamefont {Hu}, \citenamefont {Zhang},\ and\ \citenamefont
  {Shen}}]{bao14b}%
  \BibitemOpen
  \bibfield  {author} {\bibinfo {author} {\bibfnamefont {S.}~\bibnamefont
  {Bao}}, \bibinfo {author} {\bibfnamefont {J.}~\bibnamefont {Hu}}, \bibinfo
  {author} {\bibfnamefont {Z.}~\bibnamefont {Zhang}}, \ and\ \bibinfo {author}
  {\bibfnamefont {H.}~\bibnamefont {Shen}},\ }\href {\doibase
  10.1103/PhysRevC.90.045802} {\bibfield  {journal} {\bibinfo  {journal}
  {Physical Review C}\ }\textbf {\bibinfo {volume} {90}},\ \bibinfo {pages}
  {045802} (\bibinfo {year} {2014})}\BibitemShut {NoStop}%
\bibitem [{\citenamefont {Ji}\ \emph {et~al.}(2019)\citenamefont {Ji},
  \citenamefont {Hu}, \citenamefont {Bao},\ and\ \citenamefont {Shen}}]{ji19}%
  \BibitemOpen
  \bibfield  {author} {\bibinfo {author} {\bibfnamefont {F.}~\bibnamefont
  {Ji}}, \bibinfo {author} {\bibfnamefont {J.}~\bibnamefont {Hu}}, \bibinfo
  {author} {\bibfnamefont {S.}~\bibnamefont {Bao}}, \ and\ \bibinfo {author}
  {\bibfnamefont {H.}~\bibnamefont {Shen}},\ }\href {\doibase
  10.1103/PhysRevC.100.045801} {\bibfield  {journal} {\bibinfo  {journal}
  {Physical Review C}\ }\textbf {\bibinfo {volume} {100}},\ \bibinfo {pages}
  {045801} (\bibinfo {year} {2019})}\BibitemShut {NoStop}%
\bibitem [{\citenamefont {Shen}\ \emph {et~al.}(2020)\citenamefont {Shen},
  \citenamefont {Ji}, \citenamefont {Hu},\ and\ \citenamefont
  {Sumiyoshi}}]{shen20}%
  \BibitemOpen
  \bibfield  {author} {\bibinfo {author} {\bibfnamefont {H.}~\bibnamefont
  {Shen}}, \bibinfo {author} {\bibfnamefont {F.}~\bibnamefont {Ji}}, \bibinfo
  {author} {\bibfnamefont {J.}~\bibnamefont {Hu}}, \ and\ \bibinfo {author}
  {\bibfnamefont {K.}~\bibnamefont {Sumiyoshi}},\ }\href {\doibase
  10.3847/1538-4357/ab72fd} {\bibfield  {journal} {\bibinfo  {journal} {The
  Astrophysical Journal}\ }\textbf {\bibinfo {volume} {891}},\ \bibinfo {pages}
  {148} (\bibinfo {year} {2020})}\BibitemShut {NoStop}%
\bibitem [{\citenamefont {Horowitz}\ and\ \citenamefont
  {Piekarewicz}(2001)}]{horo01}%
  \BibitemOpen
  \bibfield  {author} {\bibinfo {author} {\bibfnamefont {C.~J.}\ \bibnamefont
  {Horowitz}}\ and\ \bibinfo {author} {\bibfnamefont {J.}~\bibnamefont
  {Piekarewicz}},\ }\href {\doibase 10.1103/PhysRevLett.86.5647} {\bibfield
  {journal} {\bibinfo  {journal} {Phys. Rev. Lett.}\ }\textbf {\bibinfo
  {volume} {86}},\ \bibinfo {pages} {5647} (\bibinfo {year}
  {2001})}\BibitemShut {NoStop}%
\bibitem [{\citenamefont {Bao}\ and\ \citenamefont {Shen}(2015)}]{bao15}%
  \BibitemOpen
  \bibfield  {author} {\bibinfo {author} {\bibfnamefont {S.}~\bibnamefont
  {Bao}}\ and\ \bibinfo {author} {\bibfnamefont {H.}~\bibnamefont {Shen}},\
  }\href {\doibase 10.1103/PhysRevC.91.015807} {\bibfield  {journal} {\bibinfo
  {journal} {Physical Review C}\ }\textbf {\bibinfo {volume} {91}},\ \bibinfo
  {pages} {015807} (\bibinfo {year} {2015})}\BibitemShut {NoStop}%
\bibitem [{\citenamefont {Li}\ \emph {et~al.}(2025)\citenamefont {Li},
  \citenamefont {Pang}, \citenamefont {Shen}, \citenamefont {Hu},\ and\
  \citenamefont {Sumiyoshi}}]{lisy25a}%
  \BibitemOpen
  \bibfield  {author} {\bibinfo {author} {\bibfnamefont {S.}~\bibnamefont
  {Li}}, \bibinfo {author} {\bibfnamefont {J.}~\bibnamefont {Pang}}, \bibinfo
  {author} {\bibfnamefont {H.}~\bibnamefont {Shen}}, \bibinfo {author}
  {\bibfnamefont {J.}~\bibnamefont {Hu}}, \ and\ \bibinfo {author}
  {\bibfnamefont {K.}~\bibnamefont {Sumiyoshi}},\ }\href {\doibase
  10.3847/1538-4357/ada6b3} {\bibfield  {journal} {\bibinfo  {journal} {The
  Astrophysical Journal}\ }\textbf {\bibinfo {volume} {980}},\ \bibinfo {pages}
  {54} (\bibinfo {year} {2025})}\BibitemShut {NoStop}%
\bibitem [{\citenamefont {Ji}\ \emph {et~al.}(2020)\citenamefont {Ji},
  \citenamefont {Hu}, \citenamefont {Bao},\ and\ \citenamefont {Shen}}]{ji20}%
  \BibitemOpen
  \bibfield  {author} {\bibinfo {author} {\bibfnamefont {F.}~\bibnamefont
  {Ji}}, \bibinfo {author} {\bibfnamefont {J.}~\bibnamefont {Hu}}, \bibinfo
  {author} {\bibfnamefont {S.}~\bibnamefont {Bao}}, \ and\ \bibinfo {author}
  {\bibfnamefont {H.}~\bibnamefont {Shen}},\ }\href {\doibase
  10.1103/PhysRevC.102.015806} {\bibfield  {journal} {\bibinfo  {journal}
  {Physical Review C}\ }\textbf {\bibinfo {volume} {102}},\ \bibinfo {pages}
  {015806} (\bibinfo {year} {2020})}\BibitemShut {NoStop}%
\bibitem [{\citenamefont {Ravenhall}\ \emph {et~al.}(1983)\citenamefont
  {Ravenhall}, \citenamefont {Pethick},\ and\ \citenamefont {Wilson}}]{rave83}%
  \BibitemOpen
  \bibfield  {author} {\bibinfo {author} {\bibfnamefont {D.~G.}\ \bibnamefont
  {Ravenhall}}, \bibinfo {author} {\bibfnamefont {C.~J.}\ \bibnamefont
  {Pethick}}, \ and\ \bibinfo {author} {\bibfnamefont {J.~R.}\ \bibnamefont
  {Wilson}},\ }\href {\doibase 10.1103/PhysRevLett.50.2066} {\bibfield
  {journal} {\bibinfo  {journal} {Physical Review Letters}\ }\textbf {\bibinfo
  {volume} {50}},\ \bibinfo {pages} {2066} (\bibinfo {year}
  {1983})}\BibitemShut {NoStop}%
\bibitem [{\citenamefont {Okamoto}\ \emph {et~al.}(2013)\citenamefont
  {Okamoto}, \citenamefont {Maruyama}, \citenamefont {Yabana},\ and\
  \citenamefont {Tatsumi}}]{okam13}%
  \BibitemOpen
  \bibfield  {author} {\bibinfo {author} {\bibfnamefont {M.}~\bibnamefont
  {Okamoto}}, \bibinfo {author} {\bibfnamefont {T.}~\bibnamefont {Maruyama}},
  \bibinfo {author} {\bibfnamefont {K.}~\bibnamefont {Yabana}}, \ and\ \bibinfo
  {author} {\bibfnamefont {T.}~\bibnamefont {Tatsumi}},\ }\href {\doibase
  10.1103/PhysRevC.88.025801} {\bibfield  {journal} {\bibinfo  {journal}
  {Physical Review C}\ }\textbf {\bibinfo {volume} {88}},\ \bibinfo {pages}
  {025801} (\bibinfo {year} {2013})}\BibitemShut {NoStop}%
\bibitem [{\citenamefont {Shen}\ \emph {et~al.}(2011)\citenamefont {Shen},
  \citenamefont {Toki}, \citenamefont {Oyamatsu},\ and\ \citenamefont
  {Sumiyoshi}}]{shen11}%
  \BibitemOpen
  \bibfield  {author} {\bibinfo {author} {\bibfnamefont {H.}~\bibnamefont
  {Shen}}, \bibinfo {author} {\bibfnamefont {H.}~\bibnamefont {Toki}}, \bibinfo
  {author} {\bibfnamefont {K.}~\bibnamefont {Oyamatsu}}, \ and\ \bibinfo
  {author} {\bibfnamefont {K.}~\bibnamefont {Sumiyoshi}},\ }\href {\doibase
  10.1088/0067-0049/197/2/20} {\bibfield  {journal} {\bibinfo  {journal} {The
  Astrophysical Journal Supplement Series}\ }\textbf {\bibinfo {volume}
  {197}},\ \bibinfo {pages} {20} (\bibinfo {year} {2011})}\BibitemShut
  {NoStop}%
\bibitem [{\citenamefont {Watanabe}\ \emph {et~al.}(2000)\citenamefont
  {Watanabe}, \citenamefont {Iida},\ and\ \citenamefont {Sato}}]{wata00}%
  \BibitemOpen
  \bibfield  {author} {\bibinfo {author} {\bibfnamefont {G.}~\bibnamefont
  {Watanabe}}, \bibinfo {author} {\bibfnamefont {K.}~\bibnamefont {Iida}}, \
  and\ \bibinfo {author} {\bibfnamefont {K.}~\bibnamefont {Sato}},\ }\href
  {\doibase https://doi.org/10.1016/S0375-9474(00)00197-4} {\bibfield
  {journal} {\bibinfo  {journal} {Nuclear Physics A}\ }\textbf {\bibinfo
  {volume} {676}},\ \bibinfo {pages} {455} (\bibinfo {year}
  {2000})}\BibitemShut {NoStop}%
\bibitem [{\citenamefont {Oyamatsu}\ and\ \citenamefont {Iida}(2007)}]{oyam07}%
  \BibitemOpen
  \bibfield  {author} {\bibinfo {author} {\bibfnamefont {K.}~\bibnamefont
  {Oyamatsu}}\ and\ \bibinfo {author} {\bibfnamefont {K.}~\bibnamefont
  {Iida}},\ }\href {\doibase 10.1103/PhysRevC.75.015801} {\bibfield  {journal}
  {\bibinfo  {journal} {Physical Review C}\ }\textbf {\bibinfo {volume} {75}},\
  \bibinfo {pages} {015801} (\bibinfo {year} {2007})}\BibitemShut {NoStop}%
\bibitem [{\citenamefont {Avancini}\ \emph {et~al.}(2010)\citenamefont
  {Avancini}, \citenamefont {Chiacchiera}, \citenamefont {Menezes},\ and\
  \citenamefont {Provid\^encia}}]{avan10}%
  \BibitemOpen
  \bibfield  {author} {\bibinfo {author} {\bibfnamefont {S.~S.}\ \bibnamefont
  {Avancini}}, \bibinfo {author} {\bibfnamefont {S.}~\bibnamefont
  {Chiacchiera}}, \bibinfo {author} {\bibfnamefont {D.~P.}\ \bibnamefont
  {Menezes}}, \ and\ \bibinfo {author} {\bibfnamefont {C.~m.~c.}\ \bibnamefont
  {Provid\^encia}},\ }\href {\doibase 10.1103/PhysRevC.82.055807} {\bibfield
  {journal} {\bibinfo  {journal} {Physical Review C}\ }\textbf {\bibinfo
  {volume} {82}},\ \bibinfo {pages} {055807} (\bibinfo {year}
  {2010})}\BibitemShut {NoStop}%
\bibitem [{\citenamefont {Grill}\ \emph {et~al.}(2012)\citenamefont {Grill},
  \citenamefont {Provid\^encia},\ and\ \citenamefont {Avancini}}]{gril12}%
  \BibitemOpen
  \bibfield  {author} {\bibinfo {author} {\bibfnamefont {F.}~\bibnamefont
  {Grill}}, \bibinfo {author} {\bibfnamefont {C.~m.~c.}\ \bibnamefont
  {Provid\^encia}}, \ and\ \bibinfo {author} {\bibfnamefont {S.~S.}\
  \bibnamefont {Avancini}},\ }\href {\doibase 10.1103/PhysRevC.85.055808}
  {\bibfield  {journal} {\bibinfo  {journal} {Physical Review C}\ }\textbf
  {\bibinfo {volume} {85}},\ \bibinfo {pages} {055808} (\bibinfo {year}
  {2012})}\BibitemShut {NoStop}%
\bibitem [{\citenamefont {Pais}\ \emph {et~al.}(2015)\citenamefont {Pais},
  \citenamefont {Chiacchiera},\ and\ \citenamefont {Provid\^encia}}]{pais15}%
  \BibitemOpen
  \bibfield  {author} {\bibinfo {author} {\bibfnamefont {H.}~\bibnamefont
  {Pais}}, \bibinfo {author} {\bibfnamefont {S.}~\bibnamefont {Chiacchiera}}, \
  and\ \bibinfo {author} {\bibfnamefont {C.~m.~c.}\ \bibnamefont
  {Provid\^encia}},\ }\href {\doibase 10.1103/PhysRevC.91.055801} {\bibfield
  {journal} {\bibinfo  {journal} {Physical Review C}\ }\textbf {\bibinfo
  {volume} {91}},\ \bibinfo {pages} {055801} (\bibinfo {year}
  {2015})}\BibitemShut {NoStop}%
\bibitem [{\citenamefont {Xia}\ \emph {et~al.}(2022)\citenamefont {Xia},
  \citenamefont {Maruyama}, \citenamefont {Yasutake},\ and\ \citenamefont
  {Tatsumi}}]{xia22}%
  \BibitemOpen
  \bibfield  {author} {\bibinfo {author} {\bibfnamefont {C.-J.}\ \bibnamefont
  {Xia}}, \bibinfo {author} {\bibfnamefont {T.}~\bibnamefont {Maruyama}},
  \bibinfo {author} {\bibfnamefont {N.}~\bibnamefont {Yasutake}}, \ and\
  \bibinfo {author} {\bibfnamefont {T.}~\bibnamefont {Tatsumi}},\ }\href
  {\doibase 10.1103/PhysRevD.106.063020} {\bibfield  {journal} {\bibinfo
  {journal} {Physical Review D}\ }\textbf {\bibinfo {volume} {106}},\ \bibinfo
  {pages} {063020} (\bibinfo {year} {2022})}\BibitemShut {NoStop}%
\bibitem [{\citenamefont {Schneider}\ \emph {et~al.}(2017)\citenamefont
  {Schneider}, \citenamefont {Roberts},\ and\ \citenamefont {Ott}}]{schn17}%
  \BibitemOpen
  \bibfield  {author} {\bibinfo {author} {\bibfnamefont {A.~S.}\ \bibnamefont
  {Schneider}}, \bibinfo {author} {\bibfnamefont {L.~F.}\ \bibnamefont
  {Roberts}}, \ and\ \bibinfo {author} {\bibfnamefont {C.~D.}\ \bibnamefont
  {Ott}},\ }\href {\doibase 10.1103/PhysRevC.96.065802} {\bibfield  {journal}
  {\bibinfo  {journal} {Physical Review C}\ }\textbf {\bibinfo {volume} {96}},\
  \bibinfo {pages} {065802} (\bibinfo {year} {2017})}\BibitemShut {NoStop}%
\bibitem [{\citenamefont {Bao}\ and\ \citenamefont {Shen}(2016)}]{bao16}%
  \BibitemOpen
  \bibfield  {author} {\bibinfo {author} {\bibfnamefont {S.}~\bibnamefont
  {Bao}}\ and\ \bibinfo {author} {\bibfnamefont {H.}~\bibnamefont {Shen}},\
  }\href {\doibase 10.1103/PhysRevC.93.025807} {\bibfield  {journal} {\bibinfo
  {journal} {Physical Review C}\ }\textbf {\bibinfo {volume} {93}},\ \bibinfo
  {pages} {025807} (\bibinfo {year} {2016})}\BibitemShut {NoStop}%
\bibitem [{\citenamefont {Bao}\ and\ \citenamefont {Shen}(2014)}]{bao14a}%
  \BibitemOpen
  \bibfield  {author} {\bibinfo {author} {\bibfnamefont {S.}~\bibnamefont
  {Bao}}\ and\ \bibinfo {author} {\bibfnamefont {H.}~\bibnamefont {Shen}},\
  }\href {\doibase 10.1103/PhysRevC.89.045807} {\bibfield  {journal} {\bibinfo
  {journal} {Physical Review C}\ }\textbf {\bibinfo {volume} {89}},\ \bibinfo
  {pages} {045807} (\bibinfo {year} {2014})}\BibitemShut {NoStop}%
\bibitem [{\citenamefont {Shen}\ \emph {et~al.}(1998)\citenamefont {Shen},
  \citenamefont {Toki}, \citenamefont {Oyamatsu},\ and\ \citenamefont
  {Sumiyoshi}}]{shen98b}%
  \BibitemOpen
  \bibfield  {author} {\bibinfo {author} {\bibfnamefont {H.}~\bibnamefont
  {Shen}}, \bibinfo {author} {\bibfnamefont {H.}~\bibnamefont {Toki}}, \bibinfo
  {author} {\bibfnamefont {K.}~\bibnamefont {Oyamatsu}}, \ and\ \bibinfo
  {author} {\bibfnamefont {K.}~\bibnamefont {Sumiyoshi}},\ }\href {\doibase
  10.1143/PTP.100.1013} {\bibfield  {journal} {\bibinfo  {journal} {Progress of
  Theoretical Physics}\ }\textbf {\bibinfo {volume} {100}},\ \bibinfo {pages}
  {1013} (\bibinfo {year} {1998})}\BibitemShut {NoStop}%
\bibitem [{\citenamefont {Lattimer}\ and\ \citenamefont
  {Swesty}(1991)}]{latt91}%
  \BibitemOpen
  \bibfield  {author} {\bibinfo {author} {\bibfnamefont {J.~M.}\ \bibnamefont
  {Lattimer}}\ and\ \bibinfo {author} {\bibfnamefont {F.~D.}\ \bibnamefont
  {Swesty}},\ }\href {\doibase https://doi.org/10.1016/0375-9474(91)90452-C}
  {\bibfield  {journal} {\bibinfo  {journal} {Nuclear Physics A}\ }\textbf
  {\bibinfo {volume} {535}},\ \bibinfo {pages} {331} (\bibinfo {year}
  {1991})}\BibitemShut {NoStop}%
\bibitem [{\citenamefont {M\"uller}\ and\ \citenamefont
  {Serot}(1995)}]{mull95}%
  \BibitemOpen
  \bibfield  {author} {\bibinfo {author} {\bibfnamefont {H.}~\bibnamefont
  {M\"uller}}\ and\ \bibinfo {author} {\bibfnamefont {B.~D.}\ \bibnamefont
  {Serot}},\ }\href {\doibase 10.1103/PhysRevC.52.2072} {\bibfield  {journal}
  {\bibinfo  {journal} {Physical Review C}\ }\textbf {\bibinfo {volume} {52}},\
  \bibinfo {pages} {2072} (\bibinfo {year} {1995})}\BibitemShut {NoStop}%
\bibitem [{\citenamefont {Roggero}\ \emph {et~al.}(2018)\citenamefont
  {Roggero}, \citenamefont {Margueron}, \citenamefont {Roberts},\ and\
  \citenamefont {Reddy}}]{rogg18}%
  \BibitemOpen
  \bibfield  {author} {\bibinfo {author} {\bibfnamefont {A.}~\bibnamefont
  {Roggero}}, \bibinfo {author} {\bibfnamefont {J.}~\bibnamefont {Margueron}},
  \bibinfo {author} {\bibfnamefont {L.~F.}\ \bibnamefont {Roberts}}, \ and\
  \bibinfo {author} {\bibfnamefont {S.}~\bibnamefont {Reddy}},\ }\href
  {\doibase 10.1103/PhysRevC.97.045804} {\bibfield  {journal} {\bibinfo
  {journal} {Physical Review C}\ }\textbf {\bibinfo {volume} {97}},\ \bibinfo
  {pages} {045804} (\bibinfo {year} {2018})}\BibitemShut {NoStop}%
\bibitem [{\citenamefont {Ghosh}\ \emph {et~al.}(2024)\citenamefont {Ghosh},
  \citenamefont {Shaikh}, \citenamefont {Kalita}, \citenamefont {Routaray},
  \citenamefont {Kumar},\ and\ \citenamefont {Agrawal}}]{ghos24}%
  \BibitemOpen
  \bibfield  {author} {\bibinfo {author} {\bibfnamefont {S.}~\bibnamefont
  {Ghosh}}, \bibinfo {author} {\bibfnamefont {S.}~\bibnamefont {Shaikh}},
  \bibinfo {author} {\bibfnamefont {P.~J.}\ \bibnamefont {Kalita}}, \bibinfo
  {author} {\bibfnamefont {P.}~\bibnamefont {Routaray}}, \bibinfo {author}
  {\bibfnamefont {B.}~\bibnamefont {Kumar}}, \ and\ \bibinfo {author}
  {\bibfnamefont {B.}~\bibnamefont {Agrawal}},\ }\href {\doibase
  https://doi.org/10.1016/j.nuclphysb.2024.116697} {\bibfield  {journal}
  {\bibinfo  {journal} {Nuclear Physics B}\ }\textbf {\bibinfo {volume}
  {1008}},\ \bibinfo {pages} {116697} (\bibinfo {year} {2024})}\BibitemShut
  {NoStop}%
\end{thebibliography}%

\end{document}